\documentclass[conference]{IEEEtran}

\usepackage{booktabs} 
\usepackage{tabularx}
\usepackage{xspace}
\usepackage{enumitem}
\usepackage{xcolor,colortbl}
\usepackage{amssymb}
\usepackage{amsmath}
\usepackage{multirow}
\usepackage{hyperref}
\usepackage{graphicx}
\usepackage[center]{subfigure}
\usepackage[ruled,linesnumbered]{algorithm2e} 
\usepackage[]{algorithmic} 
\usepackage{tablefootnote}
\usepackage{threeparttable}
\usepackage{makecell}
\usepackage[font={small}]{caption}

\newcommand{\subparagraph}{} 
\usepackage{titlesec} 
\titlespacing{\section}{2pt}{2pt}{2pt}
\titlespacing*{\subsection}{2pt}{2pt}{2pt}
\titlespacing*{\subsubsection}{2pt}{2pt}{2pt}

\newcolumntype{L}[1]{>{\raggedright\let\newline\\\arraybackslash\hspace{0pt}}m{#1}}
\newcolumntype{C}[1]{>{\centering\let\newline\\\arraybackslash\hspace{0pt}}m{#1}}
\newcolumntype{R}[1]{>{\raggedleft\let\newline\\\arraybackslash\hspace{0pt}}m{#1}}

\begin{document}

\title{TPA: Fast, Scalable, and Accurate Method for Approximate Random Walk with Restart on Billion Scale Graphs}

\author{\IEEEauthorblockN{Minji Yoon}
\IEEEauthorblockA{Seoul National University\\
riin55@snu.ac.kr}
\and
\IEEEauthorblockN{Jinhong Jung}
\IEEEauthorblockA{Seoul National University\\
jinhongjung@snu.ac.kr}
\and
\IEEEauthorblockN{U Kang}
\IEEEauthorblockA{Seoul National University\\
ukang@snu.ac.kr}
}

\maketitle
\newtheorem{observation}{Observation}
\newtheorem{conjecture}{Conjecture}
\newtheorem{problem}{Problem}
\newtheorem{algo}{Algorithm}
\newtheorem{definition}{Definition}
\newtheorem{lemma}{Lemma}
\newtheorem{theorem}{Theorem}
\newtheorem{property}{Property}

\newcommand{\tpa}{\textsc{TPA}\xspace}
\newcommand{\methodE}{\textsc{CPI}\xspace} 
\newcommand{\methodEIMPL}{\textsc{CPI}\xspace} 
\newcommand{\methodA}{\textsc{TPA}\xspace} 
\newcommand{\methodAN}{\textsc{TPA-NA}\xspace} 

\newcommand{\mat}[1]{\mathbf{#1}}
\newcommand{\set}[1]{\mathbf{#1}}
\newcommand{\vect}[1]{\mathbf{#1}}

\renewcommand{\r}{\vect{r}} 
\newcommand{\rst}{\r_{\text{stranger}}} %
\newcommand{\trst}{\vect{\tilde{r}}_{\text{stranger}}} %
\newcommand{\rnb}{\r_{\text{neighbor}}} %
\newcommand{\rtilde}{\vect{\tilde{r}}} 
\newcommand{\trnb}{\vect{\tilde{r}}_{\text{neighbor}}} %
\newcommand{\rfm}{\r_{\text{family}}} %

\newcommand{\p}{\vect{p}} 
\newcommand{\q}{\vect{q}} 
\newcommand{\x}{\vect{x}} 
\newcommand{\xtilde}{\vect{\tilde{x}}} 
\newcommand{\A}{\mat{A}} 
\newcommand{\NA}{\mat{\tilde{A}}} 
\newcommand{\NAT}{\mat{\tilde{A}}^{\top}} 

\newcommand{\BA}{\mat{\bar{A}}} 
\newcommand{\Bq}{\vect{\bar{q}}}
\renewcommand{\L}{\vect{f}} 

\renewcommand{\algorithmicrequire}{\textbf{Input:}}
\renewcommand{\algorithmicensure}{\textbf{Output:}}

\begin{abstract}
Given a large graph, how can we determine similarity between nodes in a fast and accurate way?
Random walk with restart (RWR) is a popular measure for this purpose and has been exploited in numerous data mining applications including ranking, anomaly detection, link prediction, and community detection.
However, previous methods for computing exact RWR require prohibitive storage sizes and computational costs, and alternative methods which avoid such costs by computing approximate RWR have limited accuracy.

In this paper, we propose \methodA, a fast, scalable, and highly accurate method for computing approximate RWR on large graphs.
\methodA exploits two important properties in RWR:
1) nodes close to a seed node are likely to be revisited in following steps due to block-wise structure of many real-world graphs, and
2) RWR scores of nodes which reside far from the seed node are proportional to their PageRank scores.
Based on these two properties, \methodA divides approximate RWR problem into two subproblems called neighbor approximation and stranger approximation.
In the neighbor approximation, \methodA estimates RWR scores of nodes close to the seed based on scores of few early steps from the seed.
In the stranger approximation, \methodA estimates RWR scores for nodes far from the seed using their PageRank.
The stranger and neighbor approximations are conducted in the preprocessing phase and the online phase, respectively.
Through extensive experiments, we show that \methodA requires up to ${\bf3.5\times}$ less time with up to ${\bf40\times}$ less memory space than other state-of-the-art methods for the preprocessing phase.
In the online phase, \methodA computes approximate RWR up to ${\bf30\times}$ faster than existing methods while maintaining high accuracy.

\end{abstract}


%
\IEEEpeerreviewmaketitle

\vspace{10pt}
\section{Introduction}
\label{sec:introduction}
Measuring similarity score between two nodes in a graph is widely recognized as a fundamental tool to analyze the graph
and has been used in various data mining tasks to gain insights about the given graph~\cite{antonellis2007query, chakrabarti2011index, fujiwara2012fast}.
Among many methods~\cite{jeh2002simrank, AxiomSimTr, lin2009matchsim} to identify similarities within graphs, random walk with restart (RWR)~\cite{pan2004automatic} has attracted considerable attention due to its ability to account for the global network structure from a particular user's point of view~\cite{he2004manifold}
and multi-faceted relationship between nodes in a graph~\cite{tong2006center}.
RWR has been widely used in various applications across different domains including ranking~\cite{DBLP:conf/icdm/JungJSK16, tong2008random}, community detection~\cite{zhu2013local, whang2013overlapping}, link prediction~\cite{backstrom2011supervised}, and anomaly detection~\cite{sun2005neighborhood}.
While RWR greatly expands its utility, it also brings a significant challenge on its computation - RWR scores are different across different seed nodes, and thus RWR needs to be recomputed for each new seed node.

To avoid enormous costs incurred by RWR computation, the majority of existing works focus on approximate RWR computation.
BRPPR~\cite{gleich2006approximating} improves RWR computation speed by limiting the amount of a Web graph data they need to access.
NB-LIN~\cite{tong2008random} computes RWR approximately by exploiting low-rank matrix approximation.
BEAR-APPROX~\cite{shin2015bear} uses a block elimination approach and precomputes several matrices including the Schur complement to exploit them in online phase.
FORA~\cite{wang2017fora} combines two methods Forward Push and Monte Carlo Random Walk with an indexing scheme.
Other methods such as FAST-PPR~\cite{lofgren2014fast} and HubPPR~\cite{wang2016hubppr} narrow down the scope of RWR problem (computing RWR scores from source to all nodes) by specifying a target node (computing a single RWR score between a source and the target node).
However, those methods are not computation-efficient enough in terms of time and memory considering the amount of their sacrificed accuracy.

\begin{figure*}[!t]
	\centering
	\includegraphics[width=.85\linewidth]{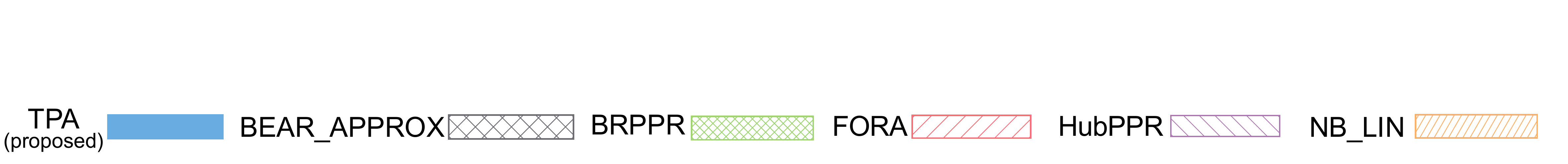}\\
	\hspace{-5mm}
	\subfigure[Size of preprocessed data]
	{
		\label{fig:perf:tpa:memory}
		\includegraphics[width=.33\linewidth]{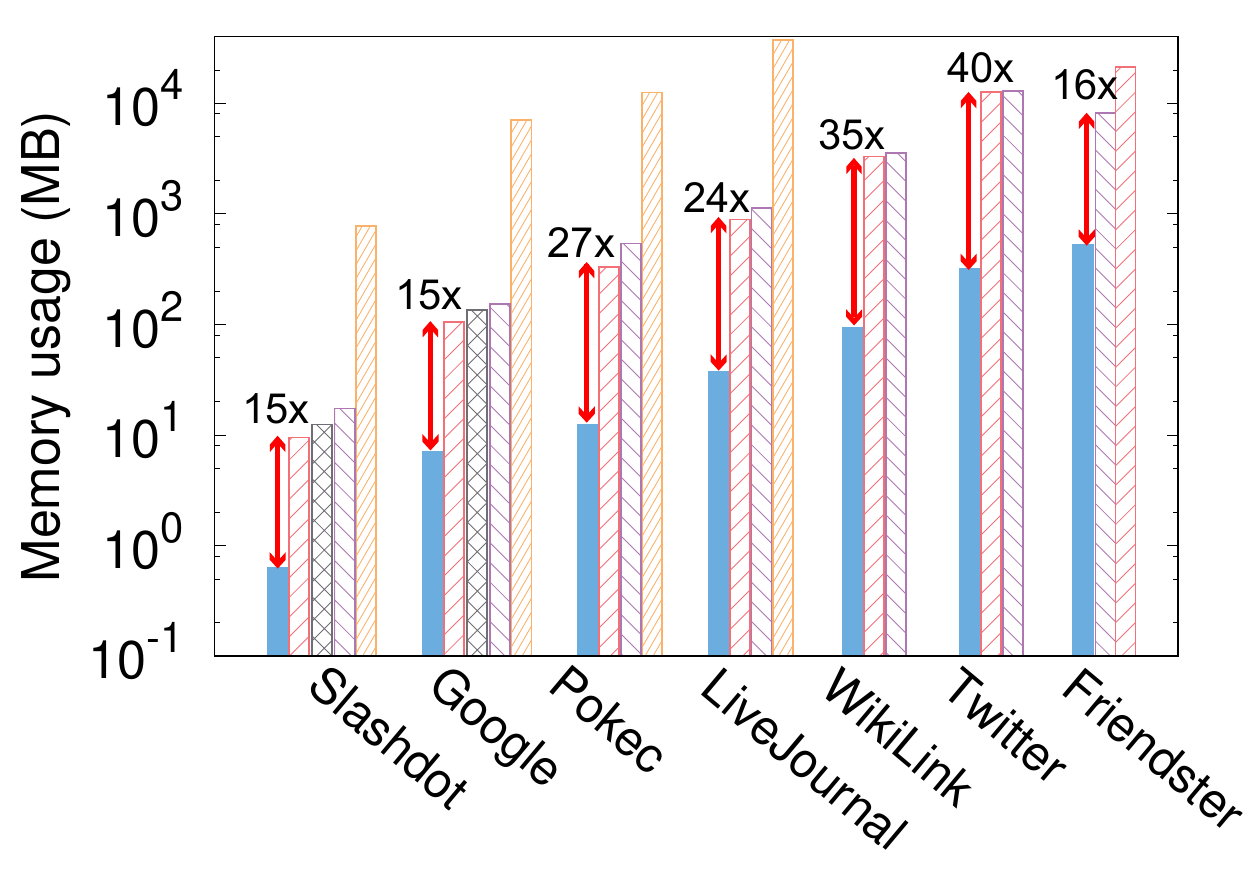}
	}
	\hspace{-5mm}
	\subfigure[Preprocessing time]
	{
		\label{fig:perf:prep_time}
		\includegraphics[width=.33\linewidth]{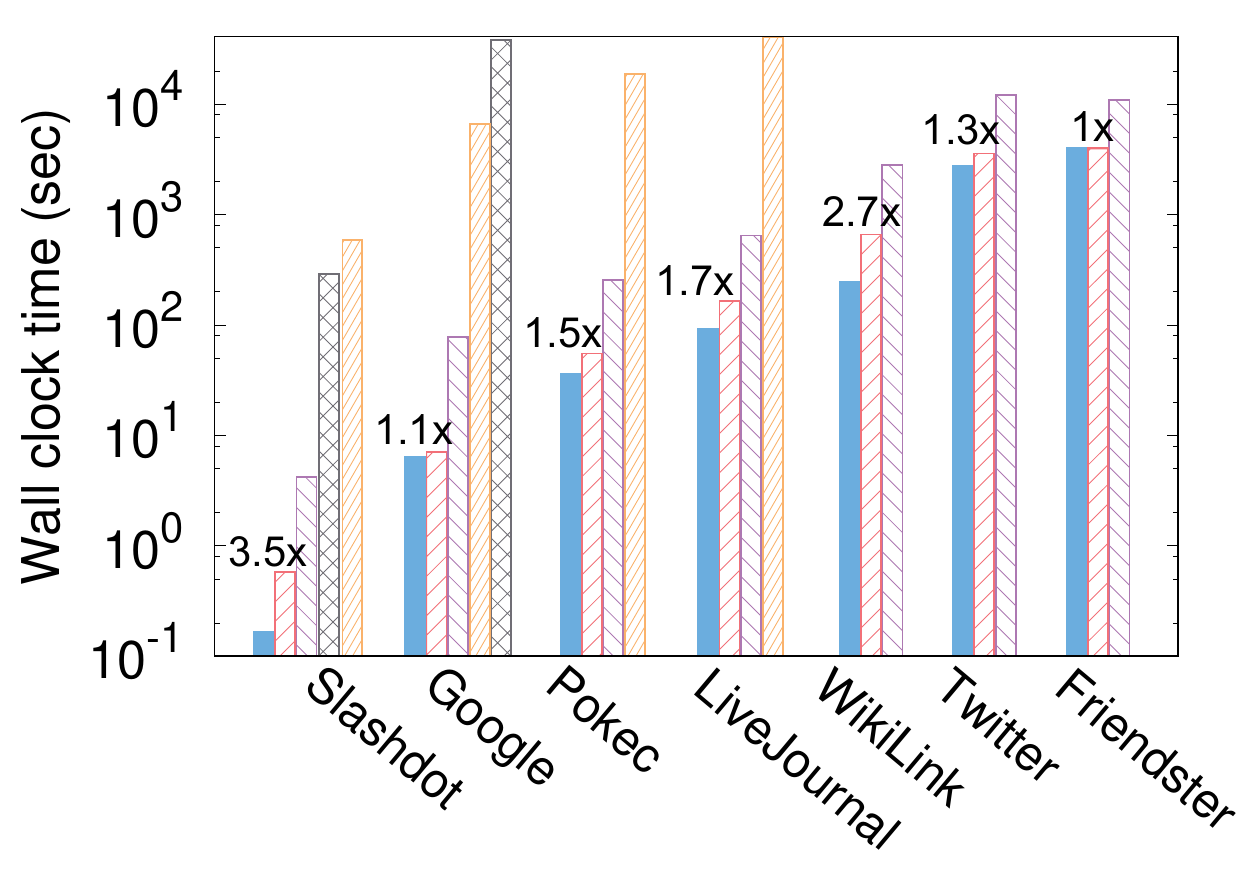}	
	}
	\hspace{-5mm}
	\subfigure[Online time]
	{
		\label{fig:perf:tpa:online_time}
		\includegraphics[width=.34\linewidth]{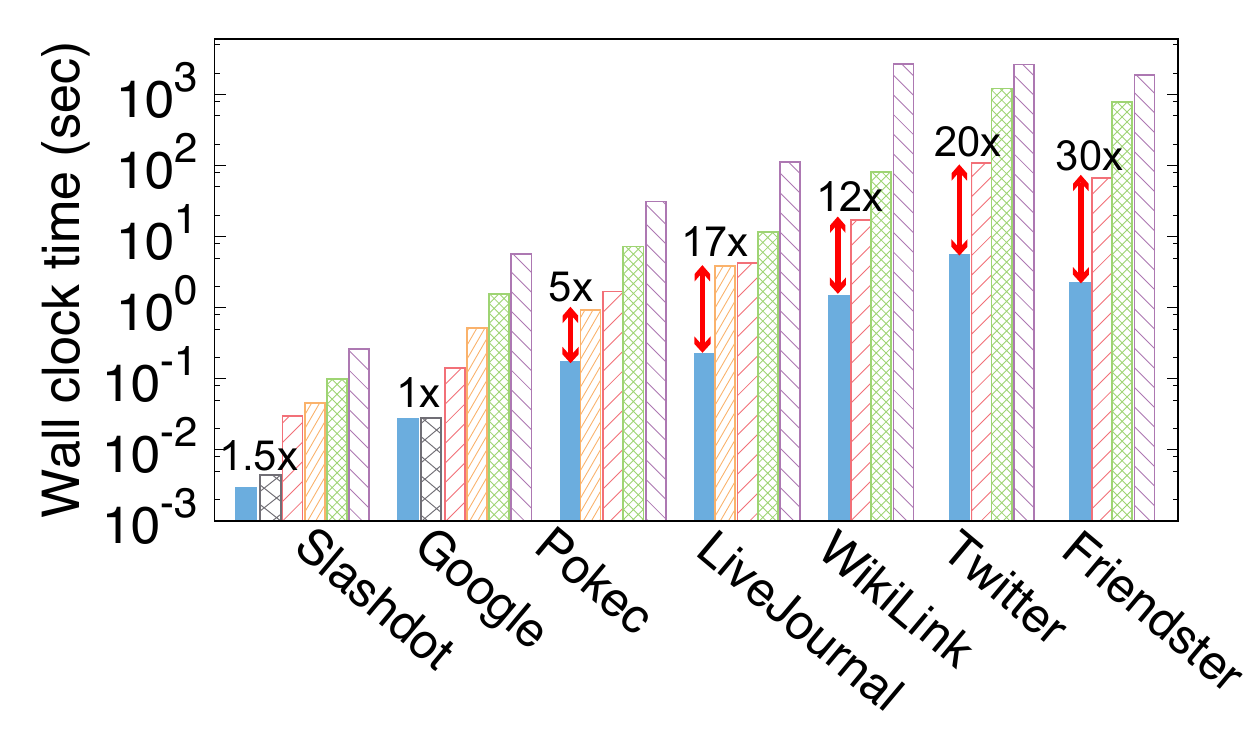}	
	}
	\hspace{-5mm}

	\caption
	{ Performance of \methodA:
		(a) compares the size of preprocessed data computed from preprocessing methods;
		(b) and (c) compare the preprocessing time and the online time, respectively, among approximate methods;
		Bars are omitted if the corresponding experiments run out of memory ($>200$GB).
		(a) \methodA uses the least amount of space for preprocessed data among preprocessing methods.
		(b) In the preprocessing phase, \methodA provides the fastest preprocessing speed among all preprocessing methods.
		(c) In the online phase, \methodA computes approximate RWR scores faster than other competitors over all datasets.
		Details on these experiments are presented in Section~\ref{sec:experiments}.
	}
	\label{fig:perf:tpa}
\end{figure*}

In this paper, we propose \methodA (Two Phase Approximation for random walk with restart), a fast, scalable, and highly accurate method for computing approximate RWR scores on billion-scale graphs.
\methodA exploits two important properties in RWR:
1) nodes close to a seed node are likely to be revisited in following steps due to block-wise structure of many real world graphs, and
2) RWR scores of nodes which reside far from the seed node are proportional to their PageRank scores. 
Based on these two properties, \methodA divides approximate RWR problem into two subproblems, the neighbor approximation and the stranger approximation.
In the neighbor approximation, \methodA estimates RWR scores of nodes close to the seed based on computation for few early steps from the seed.
In the stranger approximation, \methodA computes approximate RWR scores for nodes far from the seed using their PageRank scores.
{To divide an RWR problem into two subproblems, we use an iterative method}, cumulative power iteration (\methodE) which interprets an RWR problem as propagation of scores from a seed node across a graph.
In \methodE, $i$th iteration computes the distribution of propagated scores among nodes after $i$ steps from the seed node. 
Based on \methodE, the neighbor approximation handles iterations computed in early phase, while the stranger approximation estimates iterations computed in later phase.
The stranger and neighbor approximation phases are conducted in the preprocessing phase and the online phase, respectively.

Through extensive experiments with various real-world graphs, we demonstrate the superiority of \methodA over existing methods as shown in Figure~\ref{fig:perf:tpa}.
The main contributions of this paper are the followings:
\begin{itemize}
	\item{
		\textbf{Algorithm.}
		We propose \methodA, a fast, scalable, and highly accurate algorithm for computing approximate RWR on billion-scale graphs (Algorithms~\ref{alg:method:approx:prep}~and~\ref{alg:method:approx:online}).
		\methodA efficiently approximates RWR scores in two phases: the stranger and the neighbor approximation phases by exploiting PageRank and block-wise structure of real graphs, respectively.
	}
	\item{
		\textbf{Analysis.}
		We present an analysis of \methodA in terms of time complexity and memory requirement.
		We provide the theoretical approximation bounds for \methodA and {analyze reasons for the better approximation performance in practice than the theoretical bound suggested (Section~\ref{sec:proposed_method}).}
	}
	\item{
		\textbf{Experiment.}
		We present extensive empirical evidences for the performance of \methodA using various large real-world graphs.
		We compare \methodA with the state-of-the-art approximate RWR methods.
		Compared with other preprocessing methods, \methodA needs $3.5\times$ less time and $40\times$ less memory for the preprocessing phase.
		In the online phase, \methodA computes approximate RWR up to $30\times$ faster than other online methods, without sacrificing accuracy.
	}
\end{itemize}

The code of our method and datasets used in the paper are available at \url{http://datalab.snu.ac.kr/tpa}.
The rest of the paper is organized as follows.
In Section~\ref{sec:preliminaries}, we present preliminaries on RWR and {\methodE.}
In Section~\ref{sec:proposed_method}, we describe the proposed algorithm TPA in detail along with its theoretical analysis.
After presenting our experimental results in Section~\ref{sec:experiments}, we provide a review on related works in Section~\ref{sec:related_works} and conclude in Section~\ref{sec:conclusion}.
The symbols frequently used in this paper are summarized in Table~\ref{tab:symbols}
{and the real-world graph data used in our experiments are summarized in Table~\ref{tab:bear:dataset}.
A brief description of each dataset is in Section~\ref{sec:experiments}.}

\begin{table}[!t]
	\centering
	\small
	\caption{Table of symbols.}
	\begin{tabular}{cl}
		\toprule
		\textbf{Symbol} & \textbf{Definition} \\
		\midrule
		$G$ & {\small input graph}\\
		$n$ & {\small number of nodes in $G$}\\
		$m$ & {\small number of edges in $G$}\\
		$s$ & {\small seed node (= query node, source node)}\\
		$c$ & {\small restart probability}\\
		$\epsilon$ & {\small convergence tolerance} \\
		$\q$ & {\small ($n\times1$) seed vector}\\
		$\A$ & {\small ($n \times n$) adjacency matrix of $G$}\\
		$\NA$ & {\small ($n \times n$) row-normalized adjacency matrix of $G$}\\
		$\r_{\methodE}$ & {\small ($n\times1$) RWR vector from \methodE}\\
		$\p_{\methodE}$ & {\small ($n\times1$) PageRank vector from \methodE}\\
		$\r_{\text{\methodA}}$ &{\small ($n\times1$) approximate RWR vector} \\
											&{using neighbor and stranger approximation}\\
		$S$ & {\small ($n\times1$) starting iteration of neighbor part in \methodE}\\
		$T$ & {\small ($n\times1$) starting iteration of stranger part in \methodE}\\
		$\x^{(i)}$ & {\small ($n\times1$) interim score vector at $i$th iteration in \methodE}\\
		$\r_{\text{family}}$ & {\small ($n\times1$) sum of $\x^{(i)}$ from $0$ to $S-1$ iterations} \\
		$\r_{\text{neighbor}}$ & {\small ($n\times1$) sum of $\x^{(i)}$ from $S$ to $T-1$ iterations} \\
		$\r_{\text{stranger}}$ & {\small ($n\times1$) sum of $\x^{(i)}$ from $T$ to $\infty$ iterations} \\
		\bottomrule
	\end{tabular}
	\label{tab:symbols}
	\vspace{2mm}
\end{table}

\vspace{10pt}
\section{Preliminaries}
\label{sec:preliminaries}

In this section, we briefly review PageRank~\cite{page1999pagerank} algorithm which is used in our method for approximate value computation.
Then, we describe our target problem RWR~\cite{pan2004automatic}, and
Cumulative Power Iteration (\methodE) which computes RWR in an iterative matrix-vector multiplication form.

\subsection{PageRank}\label{sec:pagerank}
PageRank~\cite{page1999pagerank} is a widely used algorithm to measure importance of vertices in a graph.
The intuition behind PageRank is that a vertex is important if it is linked to by many important vertices.
In other words, a vertex with large number of in-edges is estimated as an important vertex with high PageRank and a vertex with few in-edges is regarded as an unimportant vertex charged with low PageRank.
PageRank scores for all nodes are represented as a PageRank score vector $\p$ which is defined by the following iterative equation:
\begin{equation*}
\p = (1-c)\NA^{\top}\p + \frac{c}{n}\vect{1}
\end{equation*}
\noindent where $\NA$ is the row-normalized adjacency matrix, $c$ is a restart probability, and $\vect{1}$ is an all-ones column vector of length $n$, the number of nodes.
If $0 < c < 1$ and $\NA$ is irreducible and aperiodic, $\p$ is guaranteed to converge to a unique solution~\cite{langville2011google}.

\subsection{Random Walk with Restart}\label{sec:rwr}
Global view of vertex importance provided by PageRank does not reflect individual preferences.
On the other hand, RWR measures each node's relevance w.r.t. a given seed node $s$ in a graph.
It assumes a random surfer who traverses the edges in the graph and occasionally restarts at node $s$.
In each step, the surfer walks along edges with probability $1-c$ or jumps to the seed node with probability $c$.
The iterative equation for an RWR score vector $\r$ is defined as follows:
\begin{equation*}
\r = (1-c)\NA^{\top}\r + c\vect{q}
\end{equation*}
\noindent where $\vect{q}$ is the seed vector with the index of the seed node $s$ set to 1 and others to 0.
In PageRank, $\vect{1}$ serves the role as a seed vector.
The only difference between a random walk in PageRank and RWR is the seed vector:
with the seed vector $\vect{1}$, a random walk in PageRank could restart from any node in the graph with uniform probability, while, with the seed vector $\vect{q}$, a random walk in RWR could restart only from the assigned seed node.

\begin{table}[!t]
	\begin{threeparttable}[t]
		\centering
		\small
		\caption{Dataset statistics: $S$ denotes the starting iteration for the neighbor approximation and $T$ denotes the starting iteration for the stranger approximation.}
		\begin{tabular}{C{15mm} | R{15mm} R{20mm} R{7mm} R{7mm}}\hline
			\toprule
			\textbf{Dataset} & \textbf{Nodes} & \textbf{Edges} & \textbf{$S$} & \textbf{$T$}\\
			\midrule
			Friendster\tnote{1}&	68,349,466&	2,586,147,869&4&20\\
			Twitter\tnote{1}&	41,652,230&	1,468,365,182&4&6\\
			WikiLink\tnote{1}&	12,150,976&	378,142,420&5&6\\
			LiveJournal\tnote{1}&	4,847,571&	68,475,391&5&10\\	
			Pokec\tnote{1}&	1,632,803&	30,622,564&5&10\\
			Google\tnote{1}&	875,713&	5,105,039&5&20\\
			Slashdot\tnote{1}&	82,144&	549,202&5&15 \\
			\bottomrule
		\end{tabular}
		\label{tab:bear:dataset}
		\begin{tablenotes}
			\item[1] {{http://konect.uni-koblenz.de/}}
			\vspace{1mm}
		\end{tablenotes}
	\end{threeparttable}
\end{table}

\subsection{\methodE: Cumulative Power Iteration}
\label{subsec:cpi}
Cumulative Power Iteration (\methodE) interprets an RWR problem as propagation of scores across a graph in an iterative matrix-vector multiplication form:
score $c$ is generated from the seed node in the beginning;
at each step, scores are divided and propagated evenly into out-edges of their current nodes with decaying coefficient $1-c$;
score $x_v$ in a node $v$ is propagated into $n_v$ out-edged neighbors of $v$ with value $\frac{1}{n_v}(1-c)x_v$.
In a matrix-vector multiplication form, $\x^{(i)}$ is an interim score vector computed from the iteration $i$ and has scores propagated across nodes at $i$th iteration as entries.
When multiplied with $(1-c)\NA^{\top}$, scores in $\x^{(i)}$ are propagated into their outgoing neighbors, and the propagated scores are stored in $\x^{(i+1)}$.
\methodE accumulates interim score vectors $\x^{(i)}$ to get the final result $\r_{\text{\methodE}}$ as follows:
\begin{align*}
	&\x^{(0)} = c\vect{q}\\
	&\x^{(i)} = (1-c)\NA^{\top}\x^{(i-1)} = c\left((1-c)\NAT\right)^{i}\q \\
	&\r_{\text{\methodE}} = \sum_{i=0}^{\infty}\vect{x}^{(i)} = c\sum_{i=0}^{\infty}\left((1-c)\NAT\right)^{i}\q
\end{align*}
\noindent We show the correctness of \methodE for RWR computation in the following Theorem~\ref{theorem:accurate_cpi}.

\begin{theorem}
	\label{theorem:accurate_cpi}
	$\r_{\text{\methodE}}$ is the true solution of the iterative equation $\r = (1-c)\NAT\r + c\q$.
	\begin{IEEEproof}
		The spectral radius of $(1-c)\NAT$ is less than one since $\NAT$ is a column stochastic matrix, which implies that $\lim_{i \rightarrow \infty}c((1-c)\NAT)^{i} \q= \mat{0}$.
		Then $\r_{\text{\methodE}}$ is convergent, and the following computation shows that $\r_{\text{\methodE}}$ obeys the steady state equation.
		\begin{align*}
		&(1-c)\NA^{\top}\r_{\text{\methodE}} + c\vect{q}\\
		&= (1-c)\NAT\left(c\sum_{i=0}^{\infty}\left((1-c)\NAT\right)^{i}\q\right)+ c\vect{q} \\
		&= c\sum_{i=1}^{\infty}\left((1-c)\NAT\right)^{i}\q + c\vect{q} \\
		&= \r_{\text{\methodE}}
		\end{align*}
	\end{IEEEproof}
\end{theorem}

\begin{algorithm} [t!]
	\small
	\begin{algorithmic}[1]
		\caption{\methodE Algorithm} \label{alg:method:exact}
		
		\REQUIRE row-normalized adjacency matrix $\NA$, seed nodes $\set{S}$, \\restart probability $c$, convergence tolerance $\epsilon$, start iteration $s_{\text{iter}}$, and terminal iteration $t_{\text{iter}}$
		\ENSURE relevance score vector $\r$
		\STATE create a seed vector $\vect{q}$ from $\set{S}$, i.e., $\vect{q}_{s} = 1/|\set{S}|$ for $s$ in $\set{S}$\\ and the others are $0$ 	\label{alg:method:exact:create_q}
		\STATE set $\r = \vect{0}$ and $\x^{(0)} = c\vect{q}$ \label{alg:method:exact:set_r}
		\FOR{iteration $i = 1$; $i \leq t_{\text{iter}}$; $i$++} \label{alg:method:exact:iter:start}
		\STATE compute $\x^{(i)} \leftarrow (1-c)(\NAT\x^{(i-1)})$ \label{alg:method:exact:compute_ith_scores}
		\IF{$i \geq s_{\text{iter}}$} \label{alg:method:exact:cumulative_r:start}
		\STATE compute $\r \leftarrow \r + \x^{(i)}$ \label{alg:method:exact:cumulative_r}
		\ENDIF \label{alg:method:exact:cumulative_r:end}
		
		\IF{$\lVert \x^{(i)} \rVert_{1} < \epsilon$} \label{alg:method:exact:check_residual:start}
		\STATE break
		\ENDIF \label{alg:method:exact:check_residual:end}
		\ENDFOR \label{alg:method:exact:iter:end}
		\BlankLine
		\RETURN $\r$
	\end{algorithmic}
\end{algorithm}

Note that there have been similar approaches~\cite{andersen2006local, jeh2003scaling} as \methodE to compute RWR,
but they do not provide any algorithm in matrix-vector multiplication form.
Thus, in this paper, we reinterpret the approaches as propagation of scores in an iterative matrix-vector multiplication form and name it \methodE.
In Algorithm~\ref{alg:method:exact}, \methodE accumulates only parts of the whole iterations using two input parameters, start iteration $s_{\text{iter}}$ and terminal iteration $t_{\text{iter}}$.
With $s_{\text{iter}}$ and $t_{\text{iter}}$, \methodE outputs the sum of $\x^{(i)}$ where $s_{\text{iter}} \leq i \leq t_{\text{iter}}$.
To get the exact RWR from \methodE, $s_{\text{iter}}$ and $t_{\text{iter}}$ are set to $0$ and $\infty$, respectively.
$s_{\text{iter}}$ and $t_{\text{iter}}$ are exploited in \methodA (Algorithms~\ref{alg:method:approx:prep} and \ref{alg:method:approx:online} in Section~\ref{sec:proposed_method}).
At first, \methodE creates a seed vector $\q$ (line 1).
For PageRank, $\q$ is set to $\frac{1}{n}\vect{1}$, and for RWR, the index of the seed node $s$ is set to 1 and others to 0 in $\q$.
In $i$th iteration, scores in $\x^{(i-1)}$ from the previous iteration ($i-1$) are propagated through $\NAT$ with decaying coefficient $1-c$ (line 4).
Then, interim score vector $\x^{(i)}$ is accumulated in RWR score vector $\r$ (line 6).
In Algorithm~\ref{alg:method:exact}, \methodE returns the sum of iterations from $s_{\text{iter}}$ to $t_{\text{iter}}$ (line 3 and 5).
Before the terminal iteration $t_{\text{iter}}$, \methodE could stop iterations when the score vector $\r$ is converged with a convergence tolerance $\epsilon$, and output $\r$ as a final score vector.
$\lVert \x^{(i)} \rVert_{1} < \epsilon$ is a condition for the final score vector $\r$ to converge (lines 8 $\sim$ 10).
\methodE could be used for PageRank and personalized PageRank which have several seed nodes.

\vspace{10pt}
\section{Proposed Method}
\label{sec:proposed_method}
\begin{figure}[!t]
	\centering
	\includegraphics[width=0.9\linewidth]{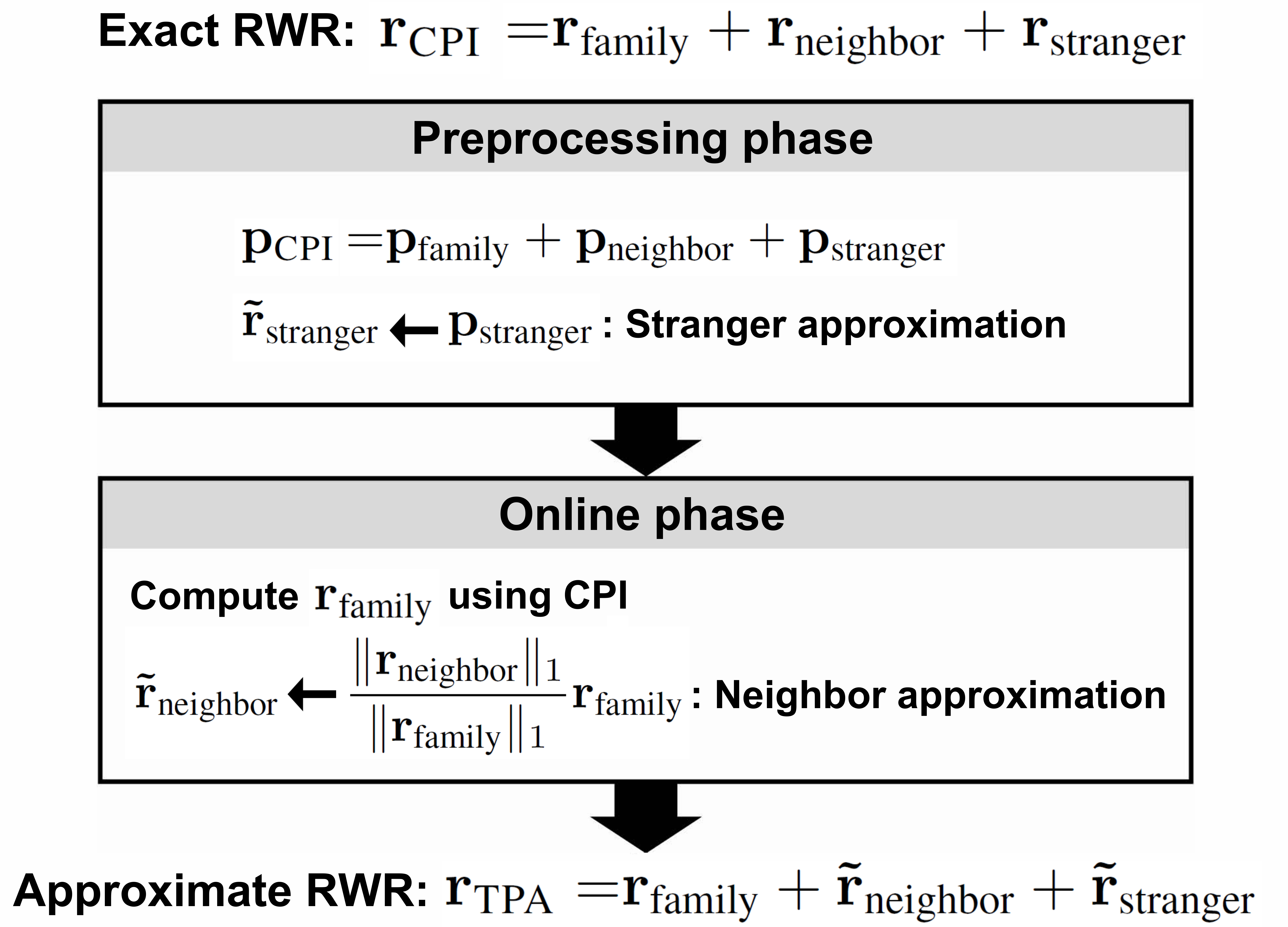}
	\caption{
		Overview: \methodA divides \methodE into three parts and applies the neighbor approximation and the stranger approximation on the neighbor part $\r_{\text{neighbor}}$ and the stranger part $\r_{\text{stranger}}$, respectively.
	}
	\label{fig:tpa:flow}
	\vspace{1mm}
\end{figure}

\methodE performs iterations until convergence (i.e., $\lVert \x^{(i)} \rVert_{1} < \epsilon$) to compute the RWR score vector $\r$.
However, considerable amount of iterations are needed for convergence and computing all the iterations is not suitable for applications which require fast RWR computation speed.
In this section, we propose \methodA which approximates RWR scores with fast speed and high accuracy.
We first divide the whole iterations in \methodE into three parts as follows:
\begin{align*}
&\r_{\text{\methodE}}\\
&= \r_{\text{family}} + \r_{\text{neighbor}} + \r_{\text{stranger}}\\
&= \underbrace{\x^{(0)}+\cdots+\x^{(S-1)}}_{\text{family part}} + \underbrace{\x^{(S)}+\cdots+\x^{(T-1)}}_{\text{neighbor part}} + \underbrace{\x^{(T)}+\cdots}_{\text{stranger part}}
\end{align*}
\noindent $S$ denotes the starting iteration in $\r_{\text{neighbor}}$, and $T$ denotes the starting iteration in $\r_{\text{stranger}}$.
The family part $\r_{\text{family}} = \x^{(0)}+\cdots+\x^{(S-1)}$ denotes the propagation of scores into nearby nodes from the seed and comprises the iterations from $0$th to ($S-1$)th in \methodE.
The neighbor part  $\r_{\text{neighbor}} = \x^{(S)}+\cdots+\x^{(T-1)}$ denotes the propagation following the family part and comprises the iterations from $S$th to ($T-1$)th.
Finally, the rest propagation part, the iterations from $T$th to the end, is denoted as $\r_{\text{stranger}} = \x^{(T)}+\cdots$.
$S$ and $T$ are tuned to give a trade-off between accuracy and computation time (more details in Section~\ref{subsec:select-t-s}).
Based on this partition, \methodA approximates the exact RWR scores $\r_{\methodE}$ by computing only $\r_{\text{family}}$ and estimating $\r_{\text{neighbor}}$ and $\r_{\text{stranger}}$.
\begin{align*}
\r_{\text{\methodA}} =\r_{\text{family}} + \rtilde_{\text{neighbor}} + \rtilde_{\text{stranger}}
\end{align*}
\methodA approximates $\r_{\text{neighbor}}$ and $\r_{\text{stranger}}$ by the neighbor approximation phase and the stranger approximation phase, respectively.
In the stranger approximation phase, \methodA estimates $\r_{\text{stranger}}$ using PageRank.
In the neighbor approximation phase, \methodA approximates $\r_{\text{neighbor}}$ using $\r_{\text{family}}$ which is the only part computed exactly.
Then the neighbor approximation and the stranger approximation are merged in the finalizing phase.
The main ideas of our proposed method are summarized as follows:

\begin{algorithm} [t!]
	\small
	\begin{algorithmic}[1]
		\caption{Preprocessing phase of \methodA}
		\label{alg:method:approx:prep}
		\REQUIRE row-normalized adjacency matrix $\NA$, restart probability $c$, convergence tolerance $\epsilon$, and starting iteration $T$ of stranger part
		\ENSURE approximate stranger score vector $\trst$
		\STATE set seeds nodes $\set{S} = \{1, \cdots, n\}$ for PageRank where $n$ is the number of nodes
		\STATE $\trst \leftarrow$ \methodE($\NA$, $\set{S}$, $c$, $\epsilon$, $T$, $\infty$) $\#$ Algorithm~\ref{alg:method:exact}
		\BlankLine
		\RETURN $\trst$
	\end{algorithmic}
\end{algorithm}

\begin{algorithm} [t!]
	\small
	\begin{algorithmic}[1]
		\caption{Online phase of \methodA}
		\label{alg:method:approx:online}
		\REQUIRE row-normalized adjacency matrix $\NA$, restart probability $c$, seed node $s$, convergence tolerance $\epsilon$, starting iteration $S$ of neighbor part, approximate stranger score vector $\trst$
		\ENSURE \methodA score vector $\r_{\methodA}$
		\STATE set a seed node $\set{S} = \{ s \}$ for RWR
		\vspace{0.6mm}
		\STATE $\r_{\text{family}} \leftarrow$ \methodE($\NA$, $\set{S}$, $c$, $\epsilon$, $0$, $S-1$) $\#$ Algorithm~\ref{alg:method:exact}
		\STATE $\rtilde_{\text{neighbor}} \leftarrow  \frac{\lVert \r_{\text{neighbor}} \rVert_{1}}{\lVert \r_{\text{family}} \rVert_{1}}\r_{\text{family}}$
		\vspace{0.6mm}
		\STATE $\r_{\text{TPA}} \leftarrow \r_{\text{family}} + \rtilde_{\text{neighbor}} + \trst$
		\BlankLine
		\RETURN $\r_{\text{TPA}}$
	\end{algorithmic}
\end{algorithm}

\begin{itemize}
	\item{
		\textbf{\methodA: stranger approximation}
		approximates the stranger part $\r_{\text{stranger}}$ in RWR with the stranger part $\p_{\text{neighbor}}$ in PageRank based on the observation
		that the distribution of scores in the stranger part is {more affected by the distribution of edges than location of a seed node}
		(Section~\ref{subsec:stranger_approx}).
	}
	\item{
		\textbf{\methodA: neighbor approximation}
		approximates the neighbor part $\r_{\text{neighbor}}$ using the family part $\r_{\text{family}}$ taking the advantage of block-wise structure of many real-world graphs (Section~\ref{subsec:neighbor_approx}).
	}
\end{itemize}

We describe each approximation phase with its accuracy analysis (Section~\ref{subsec:stranger_approx} and \ref{subsec:neighbor_approx}), and analyze time and space complexities of \methodA (Section~\ref{subsec:tpa-performance}).

\subsection{Stranger Approximation}
\label{subsec:stranger_approx}

In the stranger approximation phase, \methodA approximates the stranger part $\r_{\text{stranger}}$ using PageRank.
PageRank score vector $\p_{\text{\methodE}}$ is represented by \methodE as follows:
\begin{align*}
&\p_{\text{\methodE}} \\
&=\p_{\text{family}} + \p_{\text{neighbor}} +\p_{\text{stranger}}\\
&=\underbrace{\x'^{(0)}+\cdots+\x'^{(S-1)}}_{\text{family part}} + \underbrace{\x'^{(S)}+\cdots+\x'^{(T-1)}}_{\text{neighbor part}} + \underbrace{\x'^{(T)}+\cdots}_{\text{stranger part}}
\end{align*}
\noindent where $\x'^{(i)} =(1-c)\NAT\x'^{(i-1)}$ and $\x'^{(0)}=\frac{c}{n}\vect{1}$.
Note that the only difference between $\r_{\text{\methodE}}$ and $\p_{\text{\methodE}}$ is the seed vectors, $\x^{(0)}$ and $\x'^{(0)}$.
Then, the stranger part $\r_{\text{stranger}}$ in RWR is approximated by the stranger part $\p_{\text{stranger}}$ in PageRank.
\begin{align*}
	\rtilde_{\text{stranger}} = \p_{\text{stranger}}
\end{align*}

\textbf{Intuition.}
The amount of scores propagated into each node are determined not only by the number of in-edges of each node, but also by the distance from the seed node.
Nodes with many in-edges have many sources to receive scores, while nodes close to the seed node take in high scores since scores are decayed by factor $(1-c)$ as iteration progresses.
However, scores ($\x^{(T)},\x^{(T+1)},\cdots$) propagated in the stranger iterations are mainly determined by the number of in-edges since nodes receiving scores in the stranger iterations are already far from the seed, and thus the relative difference between their distances from the seed is too small to be considered.
Note that PageRank score vector $\p_{\methodE}$ presents the distribution of scores determined solely by the distribution of edges.
This is the main motivation for the stranger approximation: approximate the stranger part $\r_{\text{stranger}}$ in RWR with $\p_{\text{stranger}}$ in PageRank.
Since $\p_{\text{stranger}}$, the stranger part in PageRank is invariant regardless of which node is selected as a seed node, \methodA precomputes $\rtilde_{\text{stranger}}$ in the preprocessing phase (Algorithm~\ref{alg:method:approx:prep}).

\textbf{Theoretical analysis.} We show the accuracy bound of the stranger approximation in Lemma~\ref{lemma:bound_stranger}.

\begin{lemma}[Accuracy bound for $\rtilde_{\text{stranger}}$]
	\label{lemma:bound_stranger}
	Let $\r_{\text{stranger}}$ be the exact stranger part in \methodE, $\rtilde_{\text{stranger}}$ be the approximate stranger part via the stranger approximation, and $T$ be the starting iteration of the stranger part. Then $\lVert \r_{\text{stranger}} - \rtilde_{\text{stranger}} \rVert_{1} \leq 2(1-c)^{T}$.
	\begin{IEEEproof}
		$\r_{\text{stranger}}$ and $\rtilde_{\text{stranger}}$ are represented as follows:
		\begin{align*}
		\r_{\text{stranger}}&= \x^{(T)} + \x^{(T+1)} + \cdots \\
		\rtilde_{\text{stranger}} &= \x'^{(T)} + \x'^{(T+1)} + \cdots
		\end{align*}
		Then, $\lVert \r_{\text{stranger}} - \rtilde_{\text{stranger}}  \rVert_{1}$ is bounded as follows:
		\begin{align*}
		\lVert \r_{\text{stranger}} - \rtilde_{\text{stranger}}  \rVert_{1} &= \lVert (\x^{(T)} + \cdots) - (\x'^{(T)} + \cdots) \rVert_{1} \\
		&\leq \sum_{i=T}^{\infty} \lVert \x^{(i)} - \x'^{(i)} \rVert_{1}
		\end{align*}
		\noindent where the interim score vectors $\x^{(i)}$ and $\x'^{(i)}$ at $i$-th iteration in \methodE are represented as follows:
		\begin{align*}
		\x^{(i)} = (1-c)\NAT\x^{(i-1)} = c(1-c)^{i}(\NAT)^{i}\q \\
		\x'^{(i)} = (1-c)\NAT\x'^{(i-1)} = c(1-c)^{i}(\NAT)^{i}\vect{b}	
		\end{align*}
		\noindent where $\q$ is $s$-th unit vector, and $\vect{b} = \frac{1}{n}\vect{1}$.
		Suppose $(\NAT)^{i}$ is represented by $\begin{bmatrix} \vect{c}_{1}^{(i)}, \cdots, \vect{c}_{n}^{(i)} \end{bmatrix}$ where $\vect{c}_{j}^{(i)}$ is $j$-th column of the matrix $(\NAT)^{i}$, and $n$ is the number of nodes.
		Then, $\x^{(i)} - \x'^{(i)}$ is represented as follows:
		\begin{align*}
		\hspace{-1.5mm}
		\x^{(i)} - \x'^{(i)} &= c(1-c)^{i}(\NAT)^{i}(\q - \vect{b}) \\
		&= c(1-c)^{i}(\NAT)^{i}(-\frac{1}{n},-\frac{1}{n},\cdots, \underbrace{\frac{n-1}{n}}_{s\text{-th entry}},-\frac{1}{n},\cdots)^{\top}\\
		&= c(1-c)^{i}(- \frac{1}{n}\vect{c}_{1}^{(i)} \cdots +\frac{n-1}{n}\vect{c}_{s}^{(i)} \cdots - \frac{1}{n}\vect{c}_{n}^{(i)}) \\
		&= \frac{c(1-c)^{i}}{n}\sum_{j \neq s}(\vect{c}_{s}^{(i)}-\vect{c}_{j}^{(i)})
		\end{align*}
		Then, $\lVert \x^{(i)} - \x'^{(i)}\rVert_{1}$ is bounded by the following inequality:
		\begin{align*}
		\lVert \x^{(i)} - \x'^{(i)}\rVert_{1} &= \frac{c(1-c)^{i}}{n}\lVert \sum_{j \neq s}(\vect{c}_{s}^{(i)}-\vect{c}_{j}^{(i)}) \rVert_{1} \\
		&\leq \frac{c(1-c)^{i}}{n}\sum_{j \neq s}\lVert\vect{c}_{s}^{(i)}-\vect{c}_{j}^{(i)} \rVert_{1}\\
		&\leq \frac{c(1-c)^{i}}{n}\times 2(n-1) \leq 2c(1-c)^{i}
		\end{align*}
		where in the second inequality we use the fact that $\lVert \vect{c}_{j}^{(i)} \rVert_{1} = 1$ and $\lVert \vect{c}_{s}^{(i)} - \vect{c}_{j}^{(i)} \rVert_{1} \leq \lVert \vect{c}_{s}^{(i)} \rVert_{1} + \lVert \vect{c}_{j}^{(i)} \rVert_{1} = 2$, since $\NAT$ as well as $(\NAT)^{i}$ are column stochastic.
		Then, $\lVert \r_{\text{stranger}} - \rtilde_{\text{stranger}}  \rVert_{1}$ is bounded as follows:
		\begin{align*}
		\lVert \r_{\text{stranger}} - \rtilde_{\text{stranger}}  \rVert_{1} &\leq \sum_{i=T}^{\infty} \lVert \x^{(i)} - \x'^{(i)} \rVert_{1} \\
		&\leq \sum_{i=T}^{\infty}2c(1-c)^{i}
		= 2(1-c)^{T}
		\end{align*}
	\end{IEEEproof}
\end{lemma}

\begin{figure}[!t]
	\centering
	\hspace{-5mm}
	\subfigure[$\NAT$ on Slashdot]
	{
		\includegraphics[width=.45\linewidth]{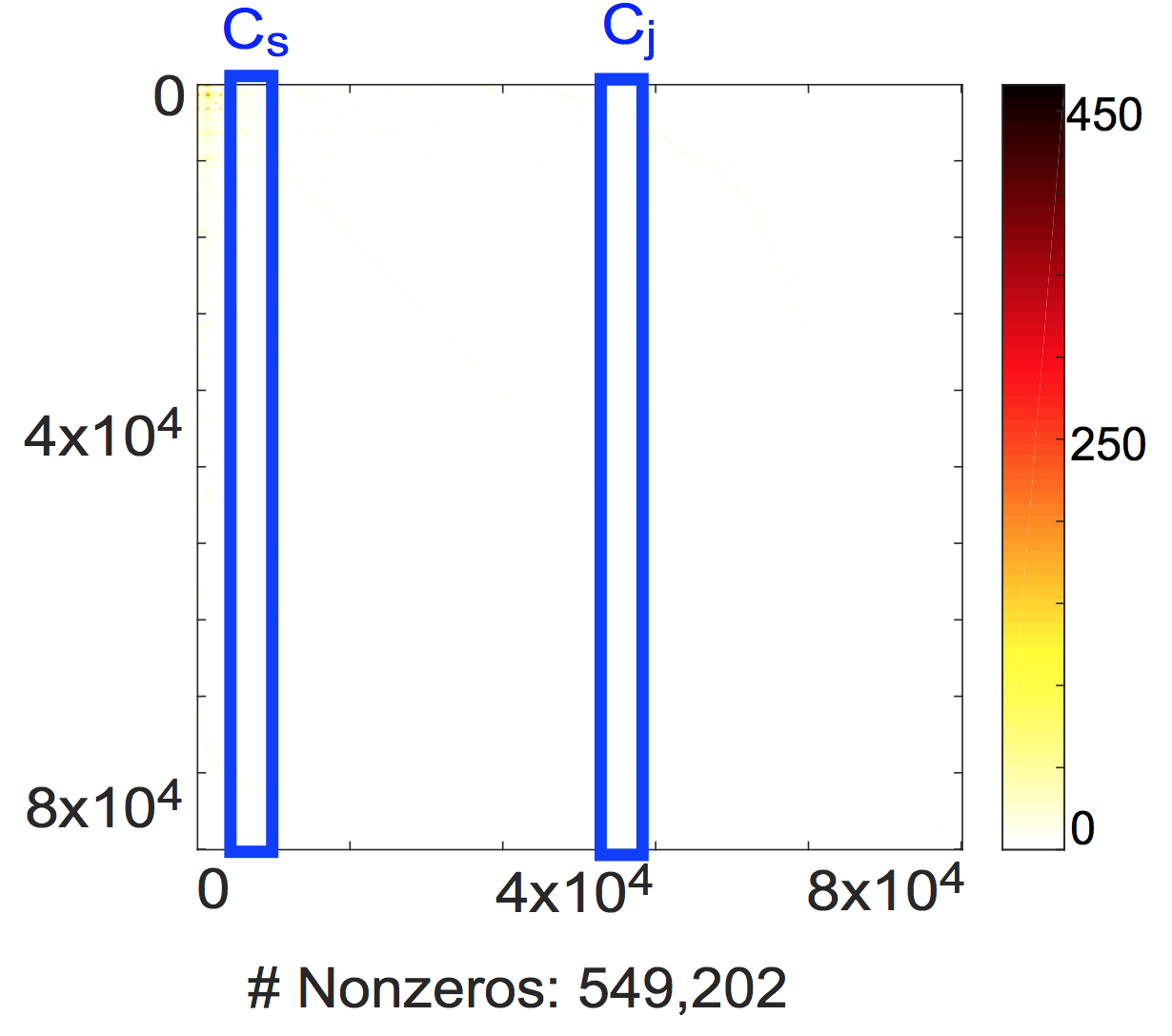}
		\vspace{-4mm}
	}
	\hspace{-5mm}
	\subfigure[$(\NAT)^{3}$ on Slashdot]
	{
		\includegraphics[width=.45\linewidth]{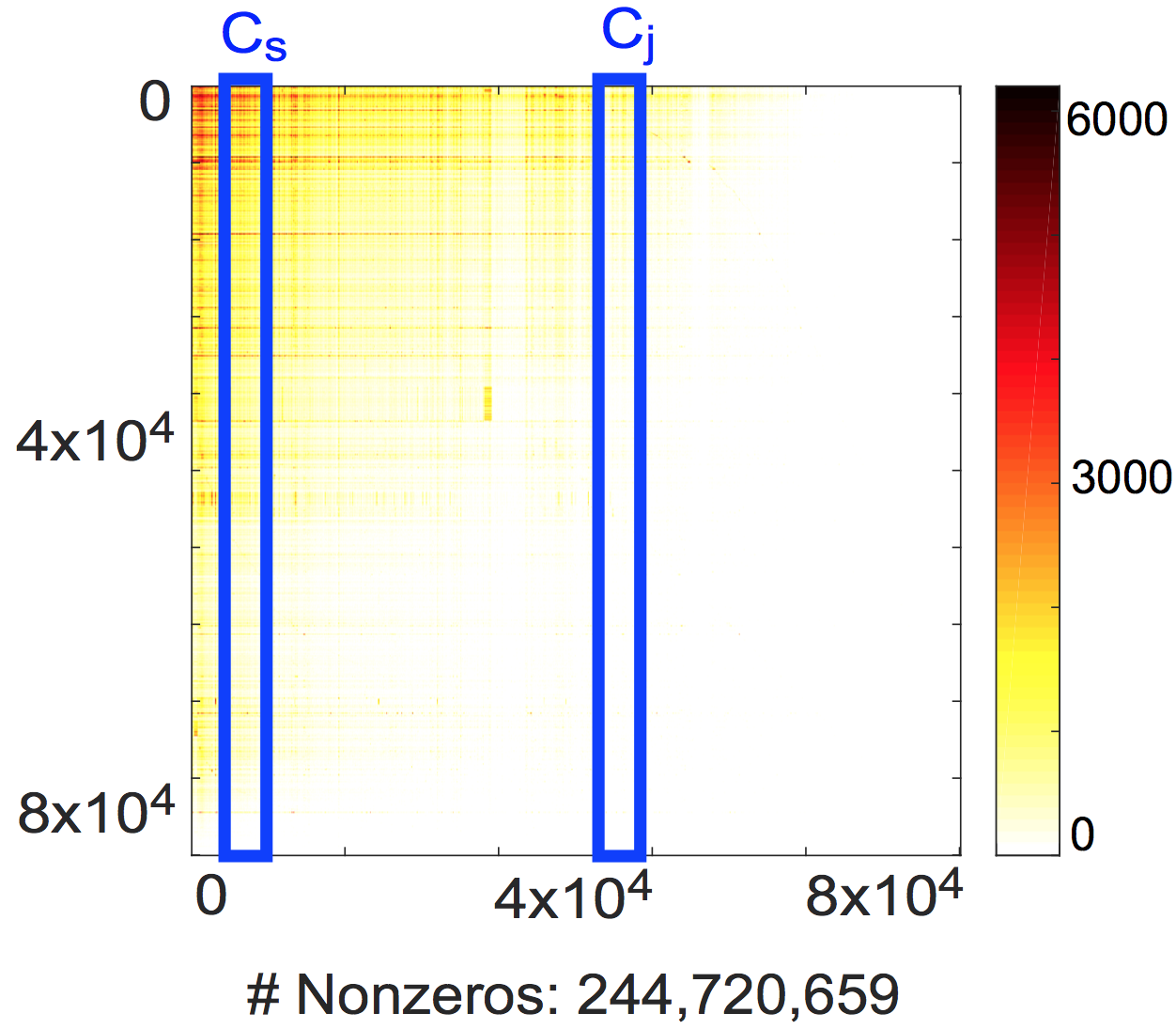}
		\vspace{-4mm}			
	}
	\hspace{-5mm}
	\\
	\vspace{-2mm}
	\hspace{-5mm}
	\subfigure[$(\NAT)^{5}$ on Slashdot]
	{
		\includegraphics[width=.45\linewidth]{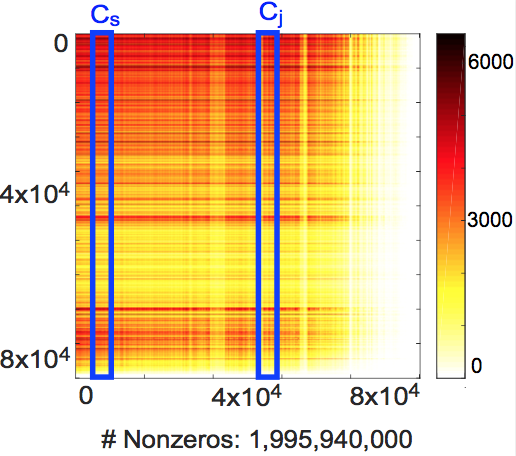}
		\vspace{-4mm}
	}
	\hspace{-5mm}
	\subfigure[$(\NAT)^{7}$ on Slashdot]
	{
		\includegraphics[width=.45\linewidth]{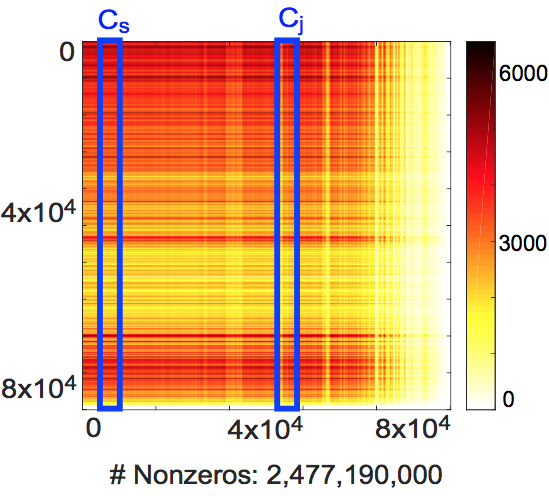}
		\vspace{-4mm}			
	}
	\vspace{-1mm}
	\caption
	{
		Distribution of nonzeros in $(\NAT)^{i}$ on Slashdot dataset:
		as $i$ increases, $(\NAT)^{i}$ has more nonzeros with denser columns $\vect{c}_{s}$ and $\vect{c}_{j}$.
		Colorbars present the number of nonzeros.
	}
	\label{fig:003:stranger-example}
\end{figure}

\begin{figure}[!t]
	\centering
	\includegraphics[width=.67\linewidth]{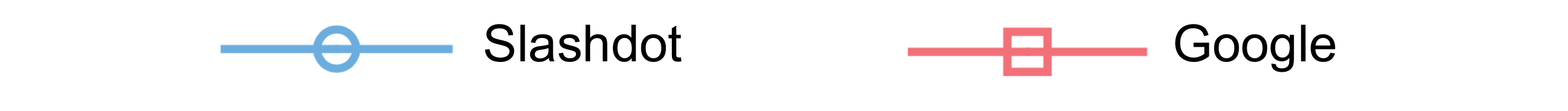}\\
	\vspace{-2mm}
	\hspace{-5mm}
	\subfigure[Nonzeros in $(\NAT)^{i}$]
	{
		\includegraphics[width=.45\linewidth]{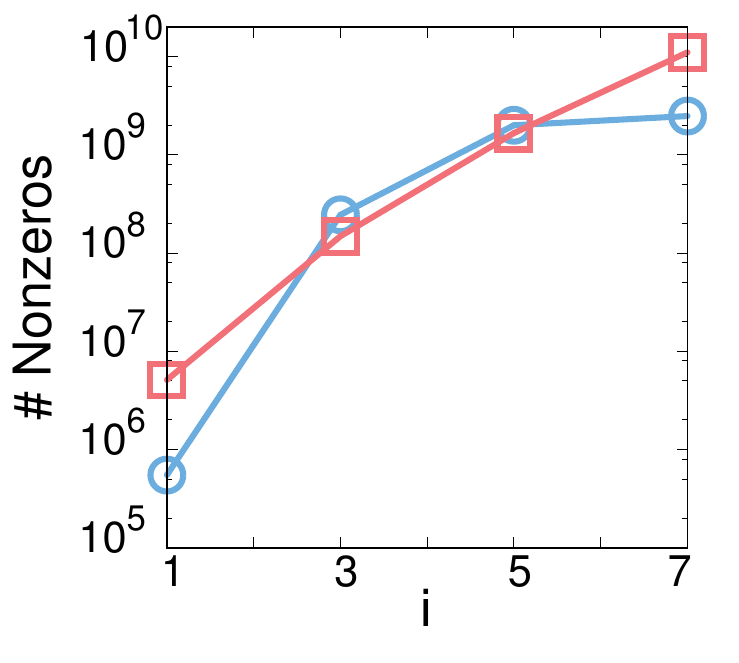}
		\vspace{-4mm}
	}
	\hspace{-2mm}
	\subfigure[$C_{i} = \frac{1}{n}\sum_{j \neq s}\lVert\vect{c}_{s}^{(i)}-\vect{c}_{j}^{(i)} \rVert_{1}$]
	{
		\includegraphics[width=.45\linewidth]{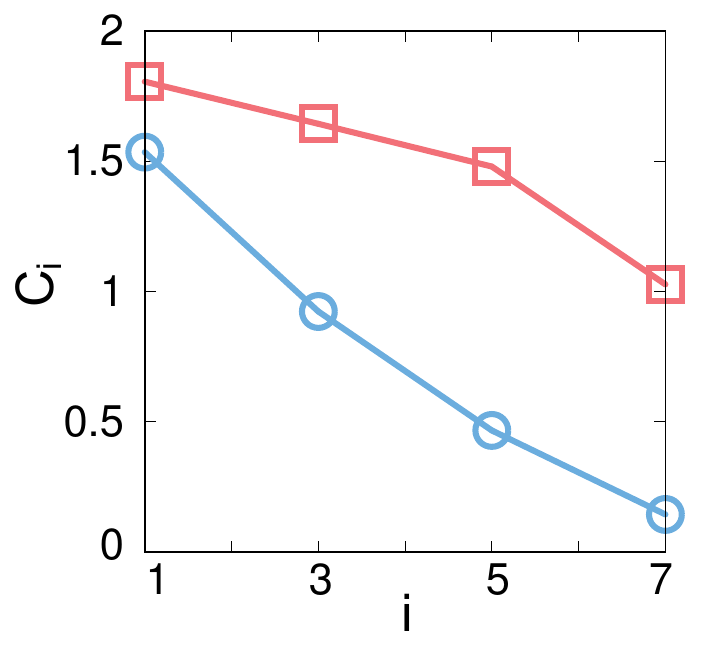}
		\vspace{-4mm}			
	}
	\vspace{-1mm}
	\caption
	{
		Number of nonzeros and  $C_{i}$ of $(\NAT)^{i}$:
		as $i$ increases, the number of nonzeros increases while $C_{i}$ decreases.
	}
	\label{fig:003:stranger}
\end{figure}

\textbf{Real-world graphs.}
From the proof of Lemma~\ref{lemma:bound_stranger}, given a seed node $s$, the $L1$ difference $\lVert \vect{c}_{s}^{(i)} - \vect{c}_{j}^{(i)} \rVert_{1}$ between column $\vect{c}_{s}^{(i)}$ and other columns $\vect{c}_{j}^{(i)}$ of a matrix $(\NAT)^{i}$ ($i = T, T+1,\cdots$) is a determining factor for accuracy of the stranger approximation.
Considering that $(\NAT)^{i}$ is a column stochastic matrix, and thus, its columns $\vect{c}_{s}^{(i)}$ and $\vect{c}_{j}^{(i)}$ are unit vectors with all non-negative entries,
$\lVert \vect{c}_{s}^{(i)} - \vect{c}_{j}^{(i)} \rVert_{1}$ becomes large (close to its maximum $2$) when $\vect{c}_{s}^{(i)}$ and $\vect{c}_{j}^{(i)}$ have their nonzero values in the different indices from each other.
Note that $\NA$ is a row-normalized adjacency matrix of a real-world graph with low density~\cite{KangF11}.
As raising the matrix $\NAT$ to the $i$th power, $(\NAT)^{i}$ tends to be a denser matrix with denser column vectors.
We present this tendency in the Slashdot dataset in Figure~\ref{fig:003:stranger-example}.
Then, the dense unit vectors $\vect{c}_{s}^{(i)}$ and $\vect{c}_{j}^{(i)}$ are likely to have nonzero values in the same indices resulting a small value of $\lVert \vect{c}_{s}^{(i)} - \vect{c}_{j}^{(i)} \rVert_{1}$.
To show this tendency in real-world graphs, we estimate the number of nonzeros in $(\NAT)^{i}$ and the average value for $C_{i} = \frac{1}{n}\sum_{j \neq s}\lVert\vect{c}_{s}^{(i)}-\vect{c}_{j}^{(i)} \rVert_{1}$ with $30$ random seeds $s$ on the Slashdot and Google datasets.
In Figure~\ref{fig:003:stranger}, as $i$ increases, the number of nonzeros increases while $C_{i}$ decreases.
This shows that the stranger approximation which approximates the stranger iterations $\x^{(i)}$ with high $i$ values ($i\geq T$) would lead to smaller errors in practice than the bound suggested in Lemma~\ref{lemma:bound_stranger}.
However, setting $T$, the starting iteration of the stranger approximation, with too high values leads to high errors in \methodA since high values for $T$ lead to high errors in the neighbor approximation.
The reasons will be discussed concretely in Section~\ref{subsec:select-t-s}.
Through extensive experiments (Section~\ref{sec:exp-error}), we present the high accuracy of the stranger approximation in real-world graphs.

\subsection{Neighbor Approximation}
\label{subsec:neighbor_approx}

In the online phase, the remaining parts $\r_{\text{family}}$ and $\r_{\text{neighbor}}$ need to be computed based on an assigned seed node.
Even though we need to compute only $T$ iterations ($\x^{(0)}, \cdots, \x^{(T-1)}$) in the online phase with the help of the stranger approximation, calculating $T$ iterations are still demanding in terms of running time.
To handle this issue, \methodA introduces the second approximation phase, the neighbor approximation.
The neighbor approximation reduces running time further by limiting computation to the family part $\r_{\text{family}}$, and estimates the neighbor part $\r_{\text{neighbor}}$ by scaling $\r_{\text{family}}$ as follows:
\begin{equation*}
	\rtilde_{\text{neighbor}} = \frac{\lVert \r_{\text{neighbor}} \rVert_{1}}{\lVert \r_{\text{family}} \rVert_{1}}\r_{\text{family}} = \frac{(1-c)^{S} - (1-c)^{T}}{1-(1-c)^{S}}\r_{\text{family}}
\end{equation*}
With restart probability $c$, the L1 norms of $\rfm$ and $\rnb$ only depend on $S$ and $T$, the starting numbers of the neighbor iterations and the stranger iterations, respectively (see Lemma~\ref{lemma:neighbor length}).

\begin{lemma}[$L1$ norms of $\r_{\text{family}}$ and $\r_{\text{neighbor}}$]
	\label{lemma:neighbor length}
	$\lVert \r_{\text{family}}\rVert_{1}$ and $\lVert \r_{\text{neighbor}}\rVert_{1}$ are $1-(1-c)^S$ and $(1-c)^S - (1-c)^T$, respectively.
	\begin{IEEEproof}
		The family part $\r_{\text{family}}$ and the neighbor part $\r_{\text{neighbor}}$ are represented as follows:
		\begin{align*}
		\r_{\text{family}}&= \x^{(0)} + \x^{(1)} + \cdots + \x^{(S-1)}\\
		\r_{\text{neighbor}} &= \x^{(S)} + \x^{(S+1)} + \cdots + \x^{(T-1)}
		\end{align*}
		where $\x^{(i)} = c(1-c)^{i}(\NAT)^{i}\q$. Then $\lVert \r_{\text{family}} \rVert_{1}$ and $\lVert \r_{\text{neighbor}} \rVert_{1}$ are represented as follows:
		\begin{align*}
		\lVert \r_{\text{family}} \rVert_{1} &= \lVert \x^{(0)} + \x^{(1)} + \cdots + \x^{(S-1)}\rVert_{1} = \sum_{i=0}^{S-1}\lVert \x^{(i)} \rVert_{1}\\
		\lVert \r_{\text{neighbor}} \rVert_{1} &= \lVert \x^{(S)} + \x^{(S+1)} + \cdots + \x^{(T-1)}\rVert_{1} = \sum_{i=S}^{T-1}\lVert \x^{(i)} \rVert_{1}
		\end{align*}
		Note that all entries of $\x^{(i)}$ are non-negative.
		Since $\NAT$ is a column stochastic matrix, $(\NAT)^{i}$ is also a column stochastic matrix.
		Hence, $\lVert (\NAT)^{i}\vect{q} \rVert_{1} =\lVert \vect{q} \rVert_{1} =1$ and $\lVert \x^{(i)} \rVert_{1}=\lVert c(1-c)^{i}(\NAT)^{i}\vect{q} \rVert_{1} = c(1-c)^{i}$.
		Then $\lVert \r_{\text{family}} \rVert_{1}$ and $\lVert \r_{\text{neighbor}} \rVert_{1}$ are written as follows:
		\begin{align*}
			\lVert \r_{\text{family}} \rVert_{1} &= \sum_{i=0}^{S-1}c(1-c)^{i} = 1 - (1-c)^S \\
			\lVert \r_{\text{neighbor}} \rVert_{1} &= \sum_{i=S}^{T-1}c(1-c)^{i} = (1-c)^{S} - (1-c)^{T}
		\end{align*}	
	\end{IEEEproof}
\end{lemma}

\noindent In the online phase, \methodA computes $\r_{\text{family}}$ at first, and estimates $\r_{\text{neighbor}}$ based on the neighbor approximation.
Finally, \methodA merges $\r_{\text{family}}$, $\rtilde_{\text{neighbor}}$ and $\rtilde_{\text{stranger}}$, and computes the approximate RWR score vector $\r_{\methodA}$ (Algorithm~\ref{alg:method:approx:online}).

\textbf{Intuition.}
In many real-world graphs, nodes inside a community are densely inter-connected to each other than to nodes in other communities.
This is an important property of real-world graphs called block-wise, community-like structure and widely exploited in graph mining~\cite{sun2005neighborhood, tong2008random}.
Our intuition for the neighbor approximation comes from this property.
Based on block-wise structure, scores started from one community are likely to be propagated into nodes in the same community repeatedly for a while.
Then we could assume that the nodes which receive scores in the early iterations (the family part) would receive scores again in the following iterations (the neighbor part).
Furthermore, the nodes which have more in-edges thus receive more scores in the early iterations would receive more scores than other nodes in the following iterations.
Note that scores propagated in the following iterations would be smaller than scores in the early iterations since scores are decayed by the decaying coefficient $(1-c)$ as iterations progress.
Based on this assumption, we maintain ratios of scores among nodes in $\r_{\text{family}}$ and scale the scores with $\frac{\lVert \r_{\text{neighbor}} \rVert_{1}}{\lVert \r_{\text{family}} \rVert_{1}}$ to reflect the smaller amount of scores in $\r_{\text{neighbor}}$.
This is the main motivation for the neighbor approximation based on block-wise structure of real-world graphs.

\textbf{Theoretical analysis.}
We show the accuracy bound for the neighbor approximation in Lemma~\ref{lemma:bound_neighbor}, and the total accuracy bound for our proposed method \methodA in Theorem~\ref{theorem:accuracy_method_approx}.

\begin{lemma}[Accuracy bound for $\rtilde_{\text{neighbor}}$]
	\label{lemma:bound_neighbor}
	Let $\rnb$ be the exact neighbor part in \methodE, and $\trnb$ be the approximate neighbor part via the neighbor approximation. Then $\lVert \rnb - \trnb \rVert_{1} \leq 2(1-c)^{S}-2(1-c)^{T}$.
	\begin{IEEEproof}
		For brevity, let $\NAT\rightarrow\BA$, $c\q\rightarrow\Bq$ and $\r_{\text{family}}=\Bq + (1-c)\BA\Bq + \cdots + ((1-c)\BA)^{S-1}\Bq \rightarrow \L$.
		We set $T = kS$ for simplicity of proof.
		Then $\r_{\text{neighbor}}$ and $\rtilde_{\text{neighbor}}$ are represented as follows:
		\begin{align*}
		\r_{\text{neighbor}} =& ((1-c)\BA)^{S}\Bq +\cdots + ((1-c)\BA)^{2S-1}\Bq\\
		&+ ((1-c)\BA)^{2S}\Bq +\cdots + ((1-c)\BA)^{3S-1}\Bq\\
		&+ \cdots \\
		&+ ((1-c)\BA)^{(k-1)S}\Bq +\cdots + ((1-c)\BA)^{kS-1}\Bq\\
		=& ((1-c)\BA)^{S}\L + \cdots + ((1-c)\BA)^{(k-1)S}\L \\
		\rtilde_{\text{neighbor}} =&  \frac{\lVert \r_{\text{neighbor}} \rVert_{1}}{\lVert \r_{\text{family}} \rVert_{1}}\r_{\text{family}} \\
		=& \frac{(1-c)^{S}-(1-c)^{T}}{1-(1-c)^{S}}\L
		\end{align*}
		Note that $\frac{(1-c)^{S}-(1-c)^{T}}{1-(1-c)^{S}}$ could be expressed as follows:
		\begin{align*}
		\frac{(1-c)^{S}-(1-c)^{T}}{1-(1-c)^{S}} &=\frac{(1-c)^{S}(1 - (1-c)^{(k-1)S})}{1-(1-c)^{S}} \\
		&= (1-c)^{S} +\cdots + (1-c)^{(k-1)S}
		\end{align*}
		Then $\rtilde_{\text{neighbor}}$ is presented as follows:
		\begin{align*}
			\rtilde_{\text{neighbor}} = (1-c)^{S}\L + \cdots + (1-c)^{(k-1)S}\L
		\end{align*}
		Then $\r_{\text{neighbor}} - \rtilde_{\text{neighbor}}$ is written as follows:
		\begin{align*}
		\small
		&\r_{\text{neighbor}} - \rtilde_{\text{neighbor}} \\
		&= (1-c)^{S}(\BA^{S}\L-\L)+(1-c)^{2S}(\BA^{2S}\L-\L)\\
		&+\cdots\\
		&+(1-c)^{(k-1)S}(\BA^{(k-1)S}\L-\L) \\
		&= \sum_{i=1}^{k-1}(1-c)^{iS}(\BA^{iS}\L -\L)
		\end{align*}
		Hence, $\lVert \r_{\text{neighbor}} - \rtilde_{\text{neighbor}} \rVert_{1}$ is bounded as follows:
		\begin{align*}
		\hspace{-3mm}
		\lVert \r_{\text{neighbor}} - \rtilde_{\text{neighbor}} \rVert_{1}
		&\leq \sum_{i=1}^{k-1}(1-c)^{iS}\lVert(\BA^{iS}\L -\L)\rVert_{1}\\
		&\leq \sum_{i=1}^{k-1}(1-c)^{iS}(\lVert\BA^{iS}\L\rVert_{1} +\lVert\L\rVert_{1})\\
		&= 2 \lVert\L\rVert_{1}\sum_{i=1}^{k-1}(1-c)^{iS}\\
		&= 2\frac{(1-c)^{S}(1-(1-c)^{(k-1)S})}{1-(1-c)^{S}}\lVert \L \rVert_{1} \\
		&= 2\frac{(1-c)^{S}-(1-c)^{kS}}{1-(1-c)^{S}}(1-(1-c)^{S})\\
		&= 2(1-c)^{S}-2(1-c)^{T}
		\end{align*}
		Note that $\lVert\BA^{iS}\L\rVert_{1}=\lVert\L\rVert_{1}$ since $\BA$ is a column stochastic matrix; thus, $\BA^{iS}$ is also a column stochastic matrix .
	\end{IEEEproof}
\end{lemma}

\begin{figure}[!t]
	\vspace{-5mm}
	\centering
	\includegraphics[width=.7\linewidth]{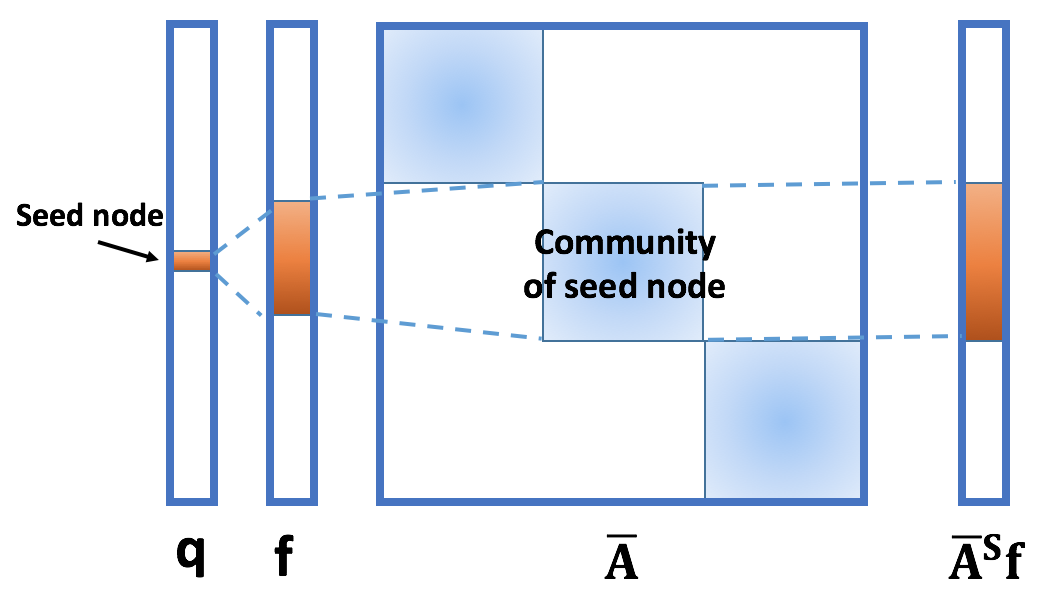}
	\vspace{-1mm}		
	\caption{
		Neighbor approximation with an ideal block-wise structure: with an ideal block-wise structure, $\L$ and $\BA^{S}\L$ have similar distributions.
	}
	\label{fig:003:neighbor-example}
\end{figure}

\begin{figure}[!t]
	\vspace{-5mm}
	\centering
	\includegraphics[width=.7\linewidth]{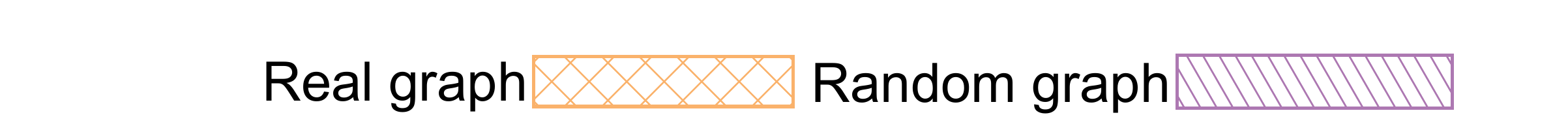}\\
	\includegraphics[width=.7\linewidth]{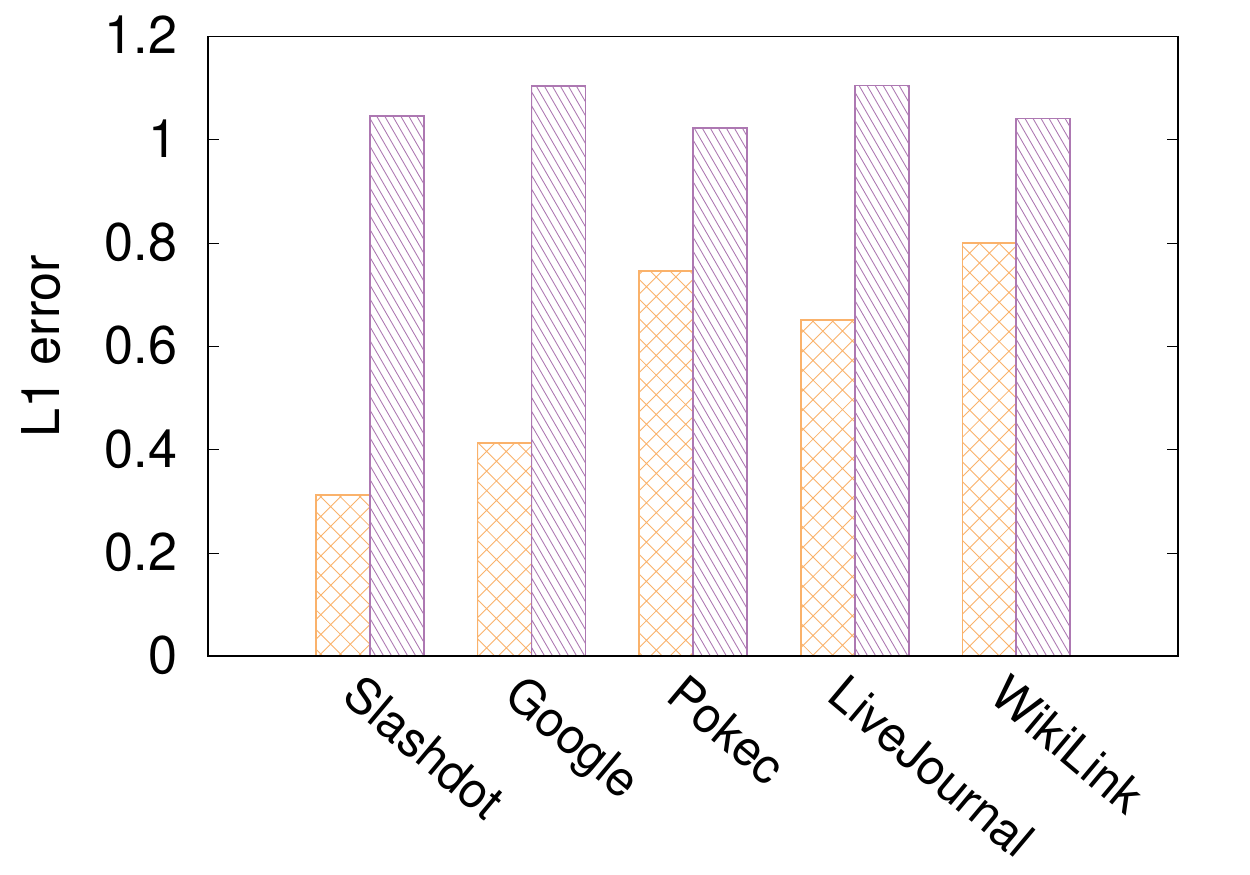}
	\vspace{-1mm}		
	\caption{
		Comparing $\lVert(\BA^{S}\L -\L)\rVert_{1}$ between real-world and random graphs:
		$\BA^{S}\L$ and $\L$ have closer values in real-world graphs with block-wise structures than those in random graphs across all datasets.
	}
	\label{fig:003:neighbor}
\end{figure}

\textbf{Real-world graphs.}
From the proof of Lemma~\ref{lemma:bound_neighbor}, $\lVert(\BA^{iS}\L -\L)\rVert_{1}$ ($i = 1, \cdots ,k-1$) is a decisive factor for the accuracy of the neighbor approximation.
$\L$ has the distribution of scores among nodes after the family iterations ($\x^{(0)},\cdots,\x^{(S-1)}$).
Multiplying $\L$ with $\BA^{S}$ means that scores in $\L$ are propagated $S$ steps further across a given graph.
As shown in Figure~\ref{fig:003:neighbor-example}, if the graph has an ideal block-wise, community-wise structure, scores in $\L$ would be mainly located in nodes around a seed node, which belong to the same community as the seed.
During the next $S$ steps, scores in $\L$ would be propagated into the nodes belonging to the same community again, without leaking into other communities.
Then, the distribution of scores in $\BA^{S}\L$ would be similar to that in $\L$.
To show that block-wise structure of real-world graphs also brings the similar effects, 
we compare $\lVert(\BA^{S}\L -\L)\rVert_{1}$ of real-world graphs (WikiLink, LiveJournal, Pokec, Google, and Slashdot) with that of random graphs.
Random graphs have the same numbers of nodes and edges as the corresponding real-world graphs, while having the random distribution of edges rather than block-wise structure.
Restart probability $c$ is set to $0.15$ and $S$ is set to $5$ for all the datasets as described in Table~\ref{tab:bear:dataset}.
As shown in Figure~\ref{fig:003:neighbor}, real-world graphs have lower $\lVert(\BA^{S}\L -\L)\rVert_{1}$ values than random graphs across all datasets.
This means that the distribution of scores ($\BA^{S}\L$) after $S$ steps is similar to the previous distribution ($\L$) in real-world graphs with the help of block-wise structure.
By this process, the neighbor approximation succeeds in achieving high accuracy for real-world graphs.

$\\$
Based on Lemmas~\ref{lemma:bound_stranger} and \ref{lemma:bound_neighbor}, we present the total accuracy bound for $\methodA$ in Theorem~\ref{theorem:accuracy_method_approx}.
Note that \methodA achieves higher accuracy in practice than the bound suggested in Theorem~\ref{theorem:accuracy_method_approx} as discussed in this section.
We show high accuracy of \methodA in real-world graphs experimentally in Section~\ref{sec:exp-error}.

\begin{theorem}[Accuracy bound for \methodA]
	\label{theorem:accuracy_method_approx}
	Let $\r_{\methodE}$ be the exact RWR score vector from \methodE, and $\r_{\methodA}$ be the approximate RWR score vector from \methodA. Then $\lVert \r_{\methodE} - \r_{\methodA} \rVert_{1}$ is bounded as follows:
	\begin{equation*}
		\lVert \r_{\methodE} - \r_{\methodA} \rVert_{1} \leq 2(1-c)^{S}
	\end{equation*}
	\begin{IEEEproof}
		Note that $\r_{\methodE}$ and $\r_{\methodA}$ are represented as follows:
		\begin{align*}
			\r_{\methodE} = \rfm + \rnb + \rst \\
			\r_{\methodA} = \rfm + \trnb + \trst
		\end{align*}
		Then $\lVert \r_{\methodE} - \r_{\methodA} \rVert_{1}$ is bounded as the following inequality:
		\begin{align*}
			\lVert \r_{\methodE} - \r_{\methodA} \rVert_{1} &= \lVert \rnb - \trnb + \rst - \trst \rVert_{1} \\
			&\leq \lVert \rnb - \trnb \rVert_{1} + \lVert \rst - \trst \rVert_{1} \\
			&\leq 2(1-c)^{T} + 2(1-c)^{S} - 2(1-c)^{T} \\
			&= 2(1-c)^{S}
		\end{align*}
		Note that $\lVert \rnb - \trnb \rVert_{1} \leq 2(1-c)^{S}-2(1-c)^{T}$ by Lemma~\ref{lemma:bound_neighbor} and $\lVert \rst - \trst \rVert_{1} \leq 2(1-c)^{T}$ by Lemma~\ref{lemma:bound_stranger}.
	\end{IEEEproof}
\end{theorem}

\noindent According to Theorem~\ref{theorem:accuracy_method_approx}, the accuracy of \methodA is bounded by $S$, the starting iteration of the neighbor approximation.
Note that $S$ also determines the scope of $\r_{\text{family}}$, thus the amount of computation needed in the online phase.
\methodA trades off the accuracy and the online computation cost using $S$.

\subsection{Selecting $S$ and $T$}
\label{subsec:select-t-s}
We set the starting iteration $S$ of the neighbor approximation considering accuracy and speed since $S$ gives a tradeoff between them.
If we set $S$ to a large value, computation time for $\r_{\text{family}}$ in the online phase escalates sharply.
Otherwise, when we set $S$ to a small value, error increases since a portion of exact computation decreases.

When we set the starting iteration $T$ of the stranger approximation to a small value, error increases sharply.
Intuitively, with small $T$, the effect of PageRank becomes higher than that of a seed node.
Theoretically, we discussed the reason with the error bound of the stranger approximation in Section~\ref{subsec:stranger_approx}.
Otherwise, when we choose a large $T$, scores of nodes far from the seed, the latter part of $\r_{\text{neighbor}}$ such as $\x^{(T-1)}$ and $\x^{(T-2)}$, are estimated by the family part $\r_{\text{family}}$ in the neighbor approximation.
Nodes far from the seed are likely to belong to different communities from that of the seed node.
Considering that the neighbor approximation assumes that nodes visited in the family part and in the neighbor part belong to the same community by block-wise structure of real-world graphs,
errors for the neighbor approximation increase significantly.
Thus we need to choose $T$ with a value which minimizes the total errors for \methodA.
We show the effects of $S$ and $T$ on the speed and accuracy of \methodA in Section~\ref{sec:exp-parameter}.

\subsection{Complexity analysis for \methodA}
\label{subsec:tpa-performance}

We analyze the time and space complexities of \methodA.
First, we evaluate the time complexity of \methodE since \methodA is based on \methodE.
\begin{lemma}[Time Complexity of \methodE]
	\label{theorem:time_cpi}
	At each iteration, \methodE takes $O(m)$ where $m$ is the number of edges in a given graph.
	In total, \methodE takes $O(m\log_{(1-c)}(\frac{\epsilon}{c}))$ time where $\log_{(1-c)}(\frac{\epsilon}{c})$ indicates the number of iterations needed for convergence.
	\begin{IEEEproof}
	\methodE computes $\x^{(i)} = (1-c)(\NA^{\top}\x^{(i-1)})$ for each iteration, and takes $O(m)$ time where $m$ is the number of nonzeros in $\NA$.
	\methodE stops the iteration with convergence when $\lVert \x^{(i)} \rVert_{1} = c(1-c)^{i} < \epsilon$.
	Then the number of iterations to be converged is $\log_{(1-c)}(\frac{\epsilon}{c})$ and the total computation time is $O(m\log_{(1-c)}(\frac{\epsilon}{c}))$.
	\end{IEEEproof}
\end{lemma}
\begin{theorem}[Time Complexity of \methodA]
	\label{theorem:time_tpa}
	\methodA takes $O(m\log_{(1-c)}(\frac{\epsilon}{c}))$ in the preprocessing phase and $O(mS)$ in the online phase where $S$ is the starting iteration of the neighbor approximation.
	\begin{IEEEproof}
		In the preprocessing phase, \methodA computes PageRank using \methodE which takes $O(m\log_{(1-c)}(\frac{\epsilon}{c}))$ time.
		In the online phase, \methodA computes $\r_{\text{family}}$ which runs $S$ iterations in \methodE; thus, it requires $O(mS)$ time.
	\end{IEEEproof}
\end{theorem}

\noindent According to Theorem~\ref{theorem:time_tpa}, the preprocessing cost and the online cost of \methodA mainly depend on the number of iterations conducted in \methodE.
Since only the family part is computed in the online phase, \methodA demands much smaller costs compared to other state-of-the-art methods as shown in Figure~\ref{fig:perf:tpa}.

\begin{theorem}[Space complexity of \methodA]
	\label{theorem:space_tpa}
	\methodA requires $O(n+m)$ memory space where $n$ and $m$ are the numbers of vertices and edges, respectively.
	\begin{IEEEproof}
		\methodA requires $O(n)$ memory space for an approximate stranger score vector $\rtilde_{\text{stranger}}$ and $O(m)$ memory space for a row-normalized adjacency matrix $\NA$.
	\end{IEEEproof}
\end{theorem}

\noindent Theorem~\ref{theorem:space_tpa} indicates that the space cost of \methodA mainly depends on $n+m$, node and edge sizes of the given graph.
As shown in Figure~\ref{fig:perf:tpa:memory}, \methodA requires reasonable memory space, and thus, succeeds in processing billion-scale graphs. 

\vspace{10pt}
\section{Experiments}
\label{sec:experiments}
\begin{figure*}[t]
	\centering
	\vspace{-3mm}
	\includegraphics[width=.8\linewidth]{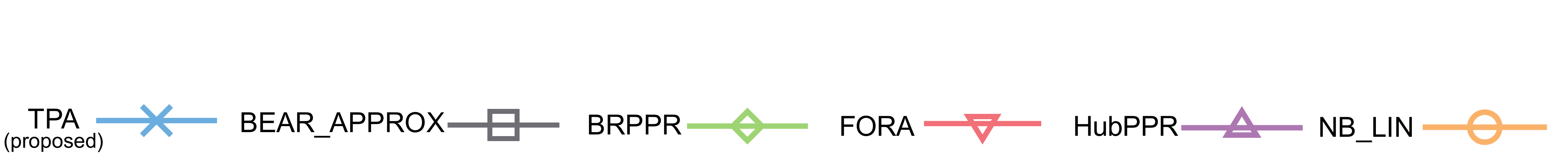}\\
	\hspace{-5mm}
	\subfigure[Slashdot]
	{
		\includegraphics[width=.25\linewidth]{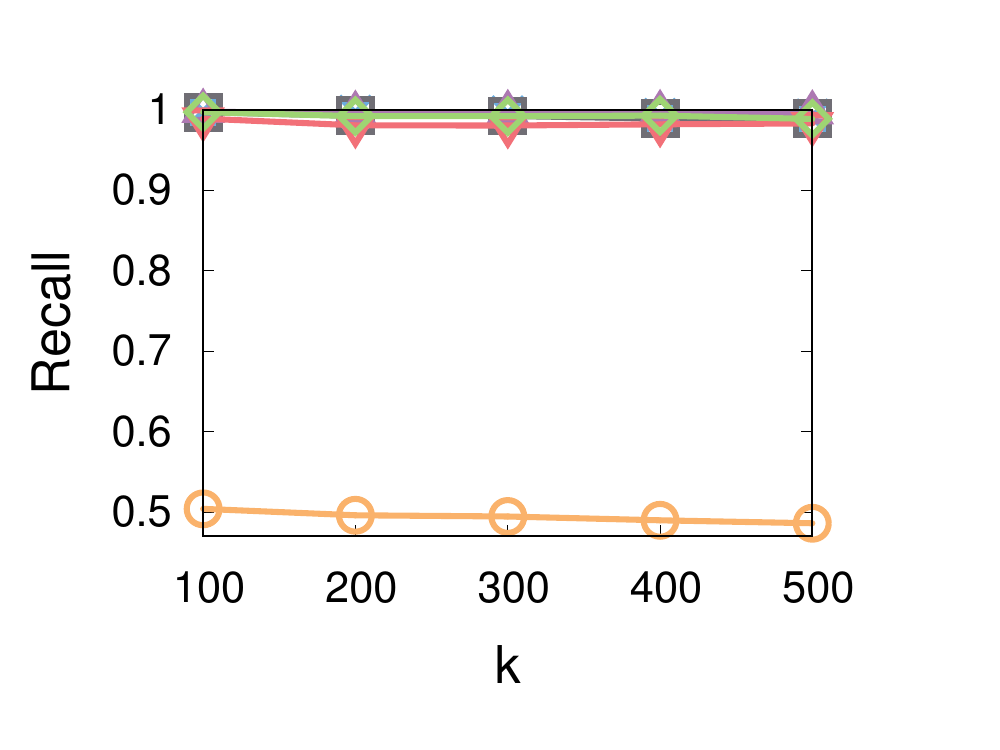}	
	}
	\hspace{-6mm}
	\subfigure[Pokec]
	{
		\includegraphics[width=.25\linewidth]{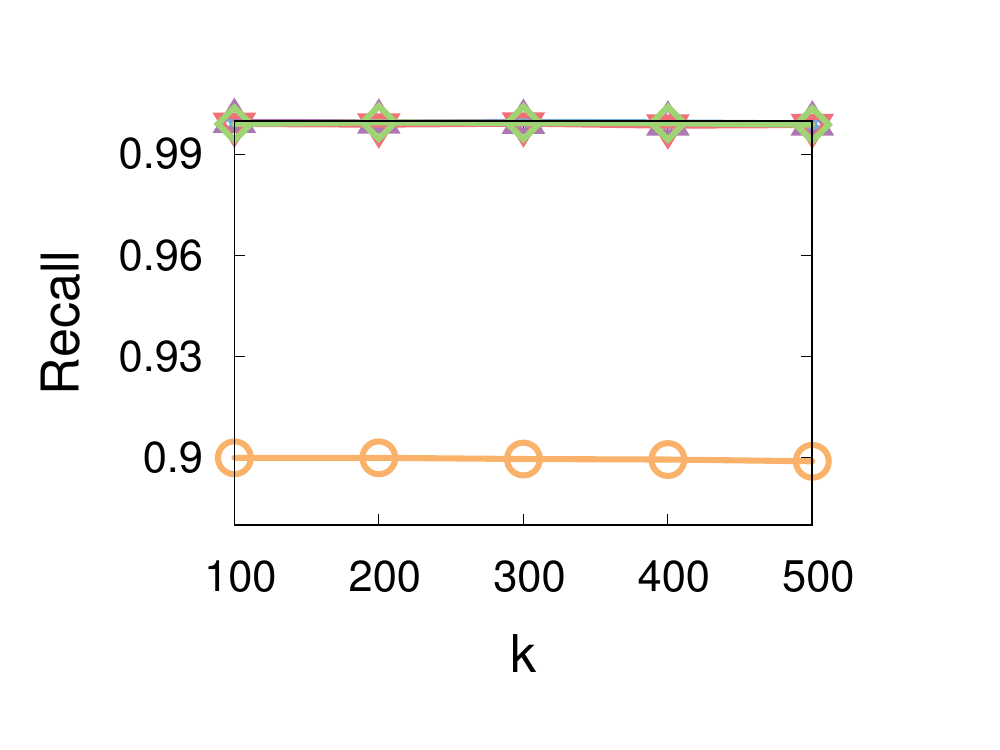}			
	}
	\hspace{-6mm}
	\subfigure[WikiLink]
	{
		\includegraphics[width=.25\linewidth]{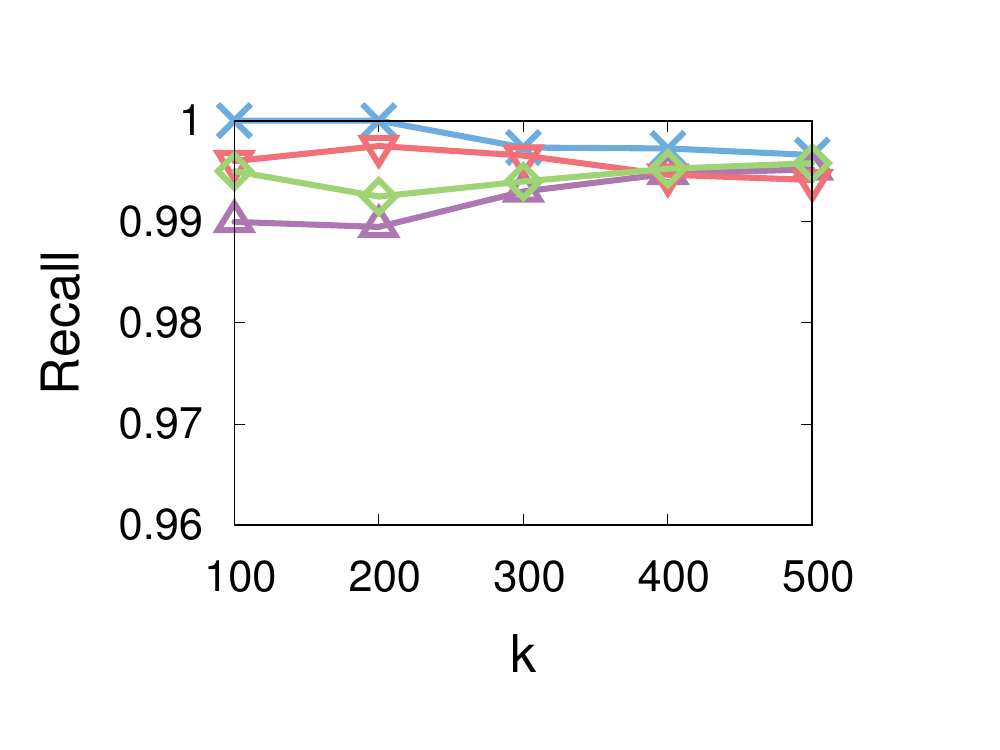}	
	}
	\hspace{-6mm}
	\subfigure[Twitter]
	{
		\includegraphics[width=.25\linewidth]{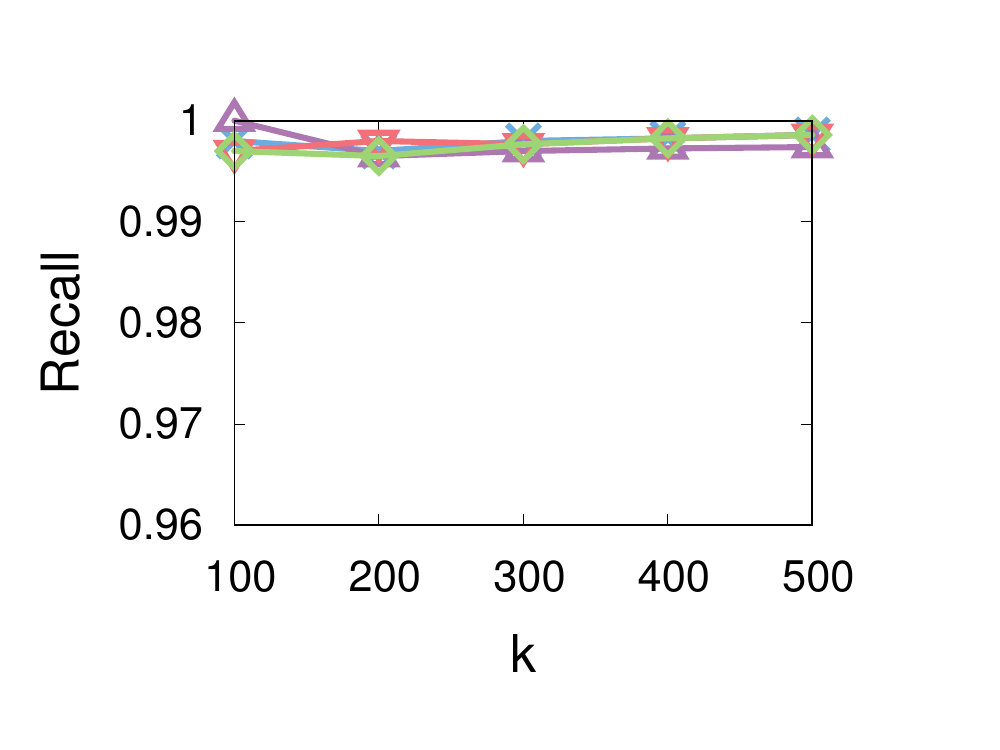}			
	}
	\hspace{-5mm}
	\caption
	{
		Recall of top-$k$ RWR vertices:
		except NB-LIN, all methods show similarly high recall on the four datasets.
		Note that \methodA requires less computation time and memory space than other methods, while maintaining the similar recall.
		Lines are omitted if corresponding experiments run out of memory ($>200$GB).
	}
	\label{fig:perf:topk}
\end{figure*}

 In this section, we experimentally evaluate the performance of \methodA compared to other approximate RWR methods.
We aim to answer the following questions:
\begin{itemize}
	{\bfseries {\item Q1 Performance of \methodA.}}
	How much does \methodA enhance the computational efficiency compared with its competitors? (Section~\ref{sec:exp-tpa})
	
	{\bfseries {\item Q2 Accuracy of \methodA.}}
	How much does \methodA reduce the approximation error in real-world graphs from the theoretical error bound? (Section~\ref{sec:exp-error})
	
	{\bfseries {\item Q3 Effects of parameters.}}
	How does the starting iteration $S$ of the neighbor approximation affect the accuracy and speed of \methodA?
	How does the starting iteration $T$ of the stranger approximation affect the accuracy of \methodA? (Section~\ref{sec:exp-parameter})
\end{itemize}

\subsection{Setup}\label{sec:exp-setup}
\subsubsection{Datasets}\label{sec:exp-dataset}
We use 7 real-world graphs to evaluate the effectiveness and efficiency of our method.
The datasets and their statistics are summarized in Table~\ref{tab:bear:dataset}.
Among them, Friendster, Twitter, LiveJournal, Pokec, and Slashdot are social networks, whereas WikiLink and Google are hyperlink networks.

\subsubsection{Environment}\label{sec:exp-environment}
All experiments are conducted on a workstation with a single core Intel(R) Xeon(R) CPU E5-2630 @ 2.2GHz and 200GB memory.
We compare \methodA with five state-of-the-art approximate RWR methods, BRPPR~\cite{gleich2006approximating}, NB-LIN~\cite{tong2008random}, BEAR-APPROX~\cite{shin2015bear}, HubPPR~\cite{wang2016hubppr}, and FORA~\cite{wang2017fora}, all of which are described in Section~\ref{sec:related_works}.
All these methods including \methodA choose an implementation showing a better performance between MATLAB and C++.
We set the restart probability $c$ to $0.15$.
The starting iteration $S$ of the neighbor approximation and the starting iteration $T$ of the stranger approximation is set differently for each graph as noted in Table~\ref{tab:bear:dataset} to gain the best performance of \methodA.
The convergence tolerance $\epsilon$ for \methodE is set to $10^{-9}$.
For each dataset, we measure the average value for $30$ random seed nodes.
To show the best performance, parameters of each competitor are set as follows:
the drop tolerance of BEAR-APPROX and NB-LIN is set to $n^{-1/2}$ and $0$, respectively;
the threshold to expand nodes in RPPR and BRPPR is set to $10^{-4}$;
parameters for the result quality guarantee of HubPPR and FORA are set to values $(1/n, 1/n, 0.5)$ as suggested in their papers~\cite{wang2016hubppr,wang2017fora}.
We obtained the source codes of HubPPR from the authors, which are optimized to compute an approximate RWR score vector.
By querying all nodes in a graph as the target nodes, HubPPR computes an approximate RWR score vector.
BePI~\cite{JungPSK17}, the state-of-the-art exact RWR method, is used to provide the ground truth RWR values in all experiments.
We compare the computational efficiency between \methodA and BePI in Appendix~\ref{appendix:bepi}.

\subsection{Competitors}\label{sec:exp-tpa}

Under the environmental settings described above, in the preprocessing phase, we estimate the preprocessing time and the size of preprocessed data of each method.
In the online phase, we estimate the computation time and the accuracy of approximate RWR scores computed from each method.
From Figures~\ref{fig:perf:tpa} to~\ref{fig:perf:topk}, \methodA runs faster than other methods, while requiring less memory space and maintaining high accuracy.

\subsubsection{Speed}\label{sec:exp-speed}
We examine the running time of \methodA in the preprocessing phase and the online phase, respectively.
Running time is measured in wall-clock time.
In the preprocessing phase, \methodA computes PageRank using \methodE to get $\rtilde_{\text{stranger}}$;
BEAR-APPROX precomputes several matrices required in the online phase;
NB-LIN computes low-rank approximation and inversion of some small size matrices;
HubPPR precomputes and indexes auxiliary information for selected hub nodes that are often involved in RWR processing;
FORA precomputes a number of random walks from nodes, and stores the destination of each walk.
As shown in Figure~\ref{fig:perf:prep_time}, \methodA performs preprocessing faster than other preprocessing methods by up to $3.5\times$.
Even though FORA shows relatively fast computation speed in the preprocessing phase, it requires up to $40\times$ larger memory space and up to $30\times$ more online computation time than \methodA.
Note that the preprocessing phase is executed only once for a graph and the online phase is executed every time for a new seed node.
Then the superior performance of \methodA for online computation has more significant effects in terms of total computation efficiency.
Under $200$GB memory capacity, BEAR-APPROX and NB-LIN fail to preprocess the datasets from Pokec and WikiLink, respectively, due to out of memory error.
In the online phase, \methodA computes an approximate RWR vector up to $30\times$ faster than other methods.
Although BEAR-APPROX takes similar online time as \methodA in the Google dataset, BEAR-APPROX takes $5923\times$ more preprocessing time than \methodA does for the same dataset.
On the contrary, \methodA maintains superior speed compared to all other methods in both phases.

\subsubsection{Memory Usage}\label{sec:exp-memory}
To compare memory efficiency, we measure how much memory each method requires for the preprocessed data.
As shown in Figure~\ref{fig:perf:tpa:memory}, compared with other preprocessing methods, \methodA requires up to $40\times$ less memory space across all the datasets.
This result shows the superior scalability of \methodA.
Under $200$GB memory capacity, BEAR-APPROX and NB-LIN consume a significant memory space, thus, are feasible only on the small datasets (LiveJournal, Pokec, Google, and Slashdot).
Although HubPPR and FORA succeed in preprocessing billion-scale graphs, they require a significant memory space for the preprocessed data.
Note that HubPPR and FORA trade off the online computation time against the size of preprocessed data~\cite{wang2016hubppr,wang2017fora}.
Thus, when they manipulate the size of preprocessed data smaller than the memory presented in Figure~\ref{fig:perf:tpa:memory}, they would require more online computation time than the one presented in Figure~\ref{fig:perf:tpa:online_time} which is already up to $30\times$ more than \methodA.

\begin{table*}[t]
	\begin{threeparttable}[t]
		\centering
		\small
		\caption
		{
				Error statistics:
				we measure $L1$ norm errors of the neighbor approximation $\rtilde_{\text{neighbor}}$, the stranger approximation $\rtilde_{\text{stranger}}$, and the final approximate RWR score vector $\r_{\text{\methodA}}$
				with regard to the exact score vectors $\r_{\text{neighbor}}$, $\r_{\text{stranger}}$, and $\r_{\text{\methodE}}$, respectively.
				Then we compare the $L1$ norm errors with their theoretical error bounds, respectively.
				The theoretical error bounds for $\rtilde_{\text{neighbor}}$, $\rtilde_{\text{stranger}}$ and $\r_{\text{\methodA}}$ are
				$2(1-c)^{S}-2(1-c)^{T}$, $2(1-c)^{T}$, and $2(1-c)^{S}$, respectively.
				Percentage denotes the ratio of $L1$ norm error in real-world graphs with regard to the theoretical error bound.
				$S$ denotes the starting iteration of the neighbor approximation and $T$ denotes the starting iteration of the stranger approximation.
				Both neighbor approximation and stranger approximation lower errors significantly from their theoretical error bounds by exploiting characteristics of real-world graphs.
		}
		\begin{tabular}{C{15mm} | R{16mm} R{10mm} R{16mm} |R{16mm}R{10mm} R{16mm} |R{16mm} R{10mm}R{16mm}}\hline
			\toprule
			\multirow{2}{*}{\textbf{Dataset}} & \multicolumn{3}{c|}{\textbf{Neighbor Approximation}} &  \multicolumn{3}{c|}{\textbf{Stranger Approximation}} &  \multicolumn{3}{c}{\textbf{\methodA}} \\ 
			& \textbf{Theoretical bound: (A)} & \textbf{Actual error: (B)} & \textbf{Percentage: (B/A)}
			& \textbf{Theoretical bound: (A)} &\textbf{Actual error: (B)} & \textbf{Percentage: (B/A)}
			& \textbf{Theoretical bound: (A)} &\textbf{Actual error: (B)} & \textbf{Percentage: (B/A)} \\
			\midrule
			Slashdot&	0.7127&	0.3367&	47.24\%&	0.1747&	0.0861&	49.27\%&	0.8874&	0.0505&	5.69\% \\
			Google&	0.8099&	0.3377&	41.70\%&	0.0775&	0.0451&	58.19\%&	0.8874&	0.1805&	20.33\%\\
			Pokec&	0.4937& 0.3041&	61.59\%&	0.3937&	0.1011&	25.68\%&	0.8874&	0.1946&	21.93\%\\
			LiveJournal&	0.4937&	0.2711&	54.91\%&	0.3937&	0.1456&	36.98\%&	0.8874&	0.2555&	 28.79\%\\
			WikiLink&	0.1331&	0.0739&	55.51\%&	0.7543&	0.2097&	27.80\%&	0.8874&	0.2370&	 26.71\%\\
			Twitter&	0.2897&	0.1953&	67.43\%&	0.7543&	0.0349&	4.63\%&	1.0440&	0.1015&	9.73\%\\
			Friendster&	0.9665&	0.4479&	46.34\%&	0.0775&	0.0419&	54.06\%&	1.0440&	0.0675&	6.46\%\\
			\bottomrule
		\end{tabular}
		\label{tab:error}
		\vspace{3mm}
	\end{threeparttable}
\end{table*}

\subsubsection{Accuracy}\label{sec:exp-accuracy}
In most applications of RWR, the typical approach is to return the top-$k$ ranked vertices of RWR vector.
For instance, in Twitter's "Who to Follow" recommendation service~\cite{gupta2013wtf}, the top-$500$ ranked users in RWR will be recommended.
Therefore, it is important to measure the accuracy of the top-$k$ results to examine the accuracy of an approximate RWR vector.
We first calculate the exact top-$k$ vertices using BePI, then evaluate the top-$k$ results of each method by measuring their recall with respect to the exact top-$k$.
For brevity, we show the result on Twitter, WikiLink, Pokec, and Slashdot; results on other graphs are similar.
As shown in Figure~\ref{fig:perf:topk}, all methods except NB-LIN provide high recall around $0.99$ across all datasets.
Note that as shown in Figure~\ref{fig:perf:tpa}, \methodA requires less computation time and smaller memory space than other methods, while maintaining the similar accuracy.

\subsection{\methodA in Real-world Graphs}
\label{sec:exp-error}
In Section~\ref{sec:proposed_method}, we analyze the error bounds of the neighbor approximation and the stranger approximation theoretically and elaborate how the approximations achieve lower errors than the theoretical bounds in real-world graphs.
The stranger approximation uses the increased density of adjacency matrices of real-world graphs as the matrices are raised to the $i$th power.
With the help of block-wise structure of real-world graphs, the neighbor approximation results in low error in practice.
In Table~\ref{tab:error}, we compare the errors of the neighbor approximation and the stranger approximation in real-world graphs with their theoretical bounds, respectively.
$S$ and $T$ used in each dataset are described in Table~\ref{tab:bear:dataset}.
The neighbor approximation lowers the error up to $42\%$ and the stranger approximation lowers the error up to $5\%$ from their theoretical error bounds, respectively.
This results show that both approximations exploit the characteristics of real-world graphs effectively.
One interesting point is that the total error of \methodA is significantly lower than the sum of errors of the neighbor approximation and the stranger approximation.
E.g., in the Slashdot dataset, the neighbor and stranger approximations lower errors to the half of the suggested theoretical bounds, but the total experimental error of \methodA decreases to $6\%$ of its theoretical upper bound.
This presents the stranger approximation and the neighbor approximation complement each other effectively.
The neighbor approximation could not consider nodes which are not visited in the family iterations since it approximates the neighbor iterations only based on the family iterations.
On the other hand, the stranger approximation could not consider the effect of seed node since it precomputes PageRank in the preprocessing phase without any information about which node would be a seed node.
Merged with the stranger approximation, the neighbor approximation acquires information about nodes across whole graphs, stored in PageRank.
On the other hand, merged with the neighbor approximation, the stranger approximation has a chance to put more priority on the seed node.
\methodA compensates the weak points of each approximations successfully.

\subsection{Effects of Parameters}\label{sec:exp-parameter}

\begin{figure}[!t]
	\hspace{3mm}
	\vspace{-2mm}
	\includegraphics[width=.8\linewidth]{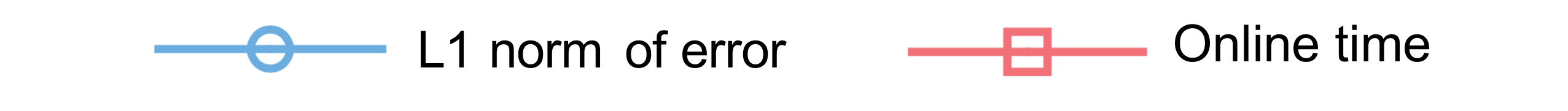}\\
	\centering
	\hspace{-6mm}
	\subfigure[Online time vs. L1 norm of error on the LiveJournal dataset]
	{
		\includegraphics[width=.46\linewidth]{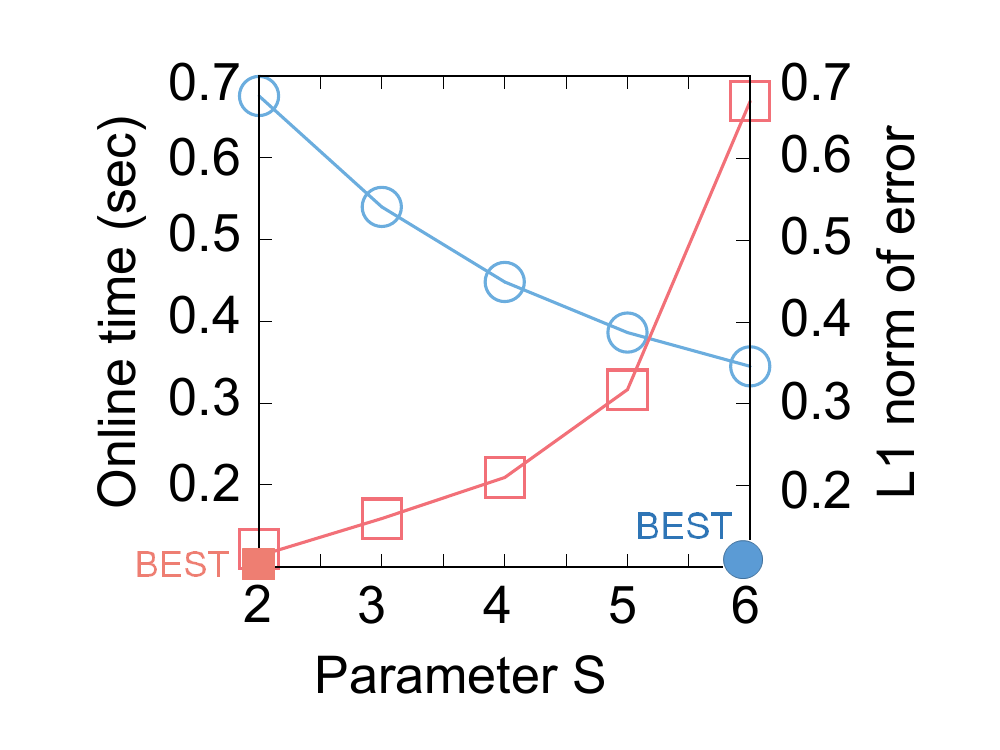}	
	}
	\hspace{-0.5mm}
	\subfigure[Online time vs. L1 norm of error on the Pokec dataset]
	{
		\includegraphics[width=.46\linewidth]{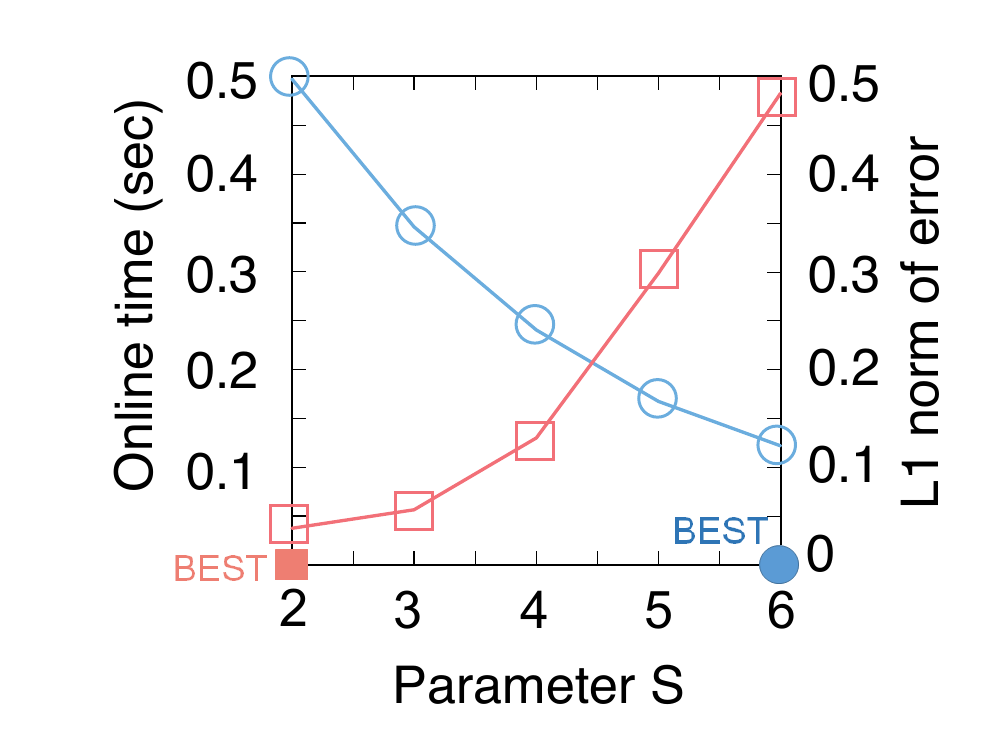}			
	}
	\caption
	{
		Effects of $S$: with small $S$, \methodA runs fast with high $L1$ norm error.
		On the other hand, with large $S$, \methodA takes long computation time with low $L1$ error.
	}
	\label{fig:perf:s}
	\vspace{1mm}
	\includegraphics[width=.9\linewidth]{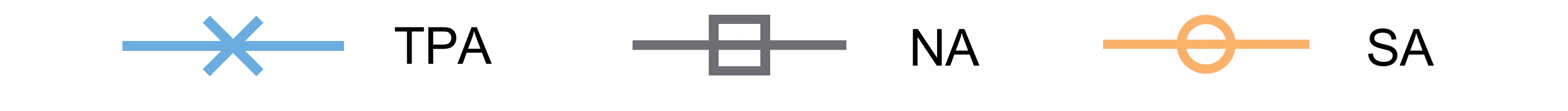}\\
	\vspace{-4mm}
	\hspace{-6mm}
	\subfigure[L1 norm of error on the LiveJournal dataset]
	{
		\includegraphics[width=.45\linewidth]{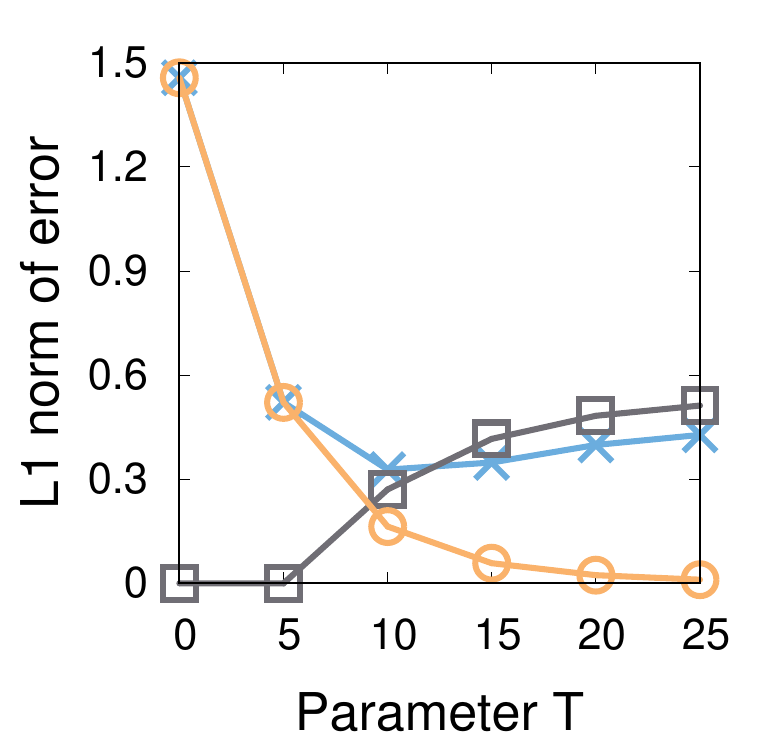}	
	}
	\hspace{-3mm}
	\subfigure[L1 norm of error on the Pokec dataset]
	{
		\includegraphics[width=.45\linewidth]{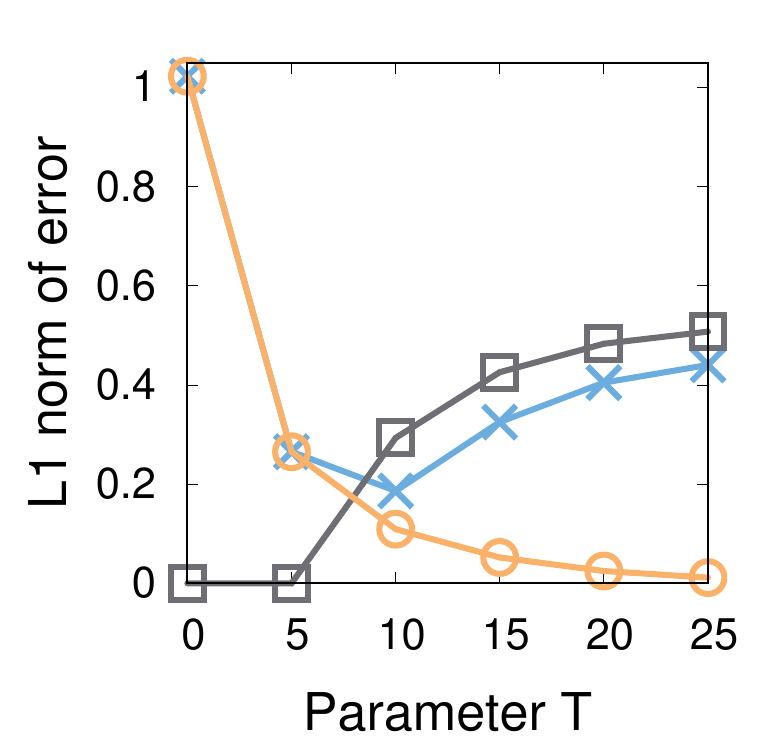}			
	}
	\caption
	{
		Effects of $T$: as $T$ increases,
		$L1$ norm error of the neighbor approximation (NA) increases, that of the stranger approximation (SA) decreases, and that of \methodA decreases at first and then rebounds from  $T=10$.
	}
	\label{fig:perf:t}
\end{figure}

We discuss the effects of two parameters $S$ and $T$ in this subsection.
We first investigate the effects of $S$, the starting iteration of the neighbor approximation, on the performance of \methodA.
We measure online computation time and $L1$ norm error of \methodA varying $S$.
During this experiment, $T$ is fixed to $10$.
As shown in Figure~\ref{fig:perf:s}, as $S$ increases, online time increases sharply while $L1$ norm error decreases since a portion of the exact computation $\r_{\text{family}}$ increases.
Thus, $S$ is selected to a proper number considering the tradeoff between accuracy and running time of \methodA.

Next, we examine the effects of $T$, the starting iteration of the stranger approximation, on the accuracy of \methodA.
We measure $L1$ norm errors of the neighbor approximation, the stranger approximation, and \methodA, respectively, varying $T$.
Note that $S$ is fixed to $5$ during this experiment.
In Figure~\ref{fig:perf:t}, as $T$ increases, $L1$ error of the neighbor approximation increases, that of the stranger approximation decreases, and that of \methodA decreases at first and then rebounds from $T=10$.
With small $T$, the stranger approximation applies to nodes close to the seed, then, the nodes are estimated by their PageRank scores and the effects of their close distances from the seed are ignored.
This leads to high $L1$ norm error of the stranger approximation.
On the other hand, with large $T$, the neighbor approximation applies to nodes far from the seed.
The nodes which reside far from the seed are likely to belong to different communities from that of the seed.
However, by the neighbor approximation, such nodes are estimated as the same community members as the seed.
Then, $L1$ norm error of the neighbor approximation becomes high.
Thus $T$ is set to a value which minimizes the total $L1$ norm error of \methodA.

\vspace{10pt}
\section{Related Works}
\label{sec:related_works}
In this section, we review previous approximate methods for RWR.
To avoid enormous costs incurred by RWR computation, many efforts have been devoted to estimating RWR in a cost-efficient way while sacrificing little accuracy.
Gleich et al.~\cite{gleich2006approximating} introduced boundary restricted personalized PageRank (BRPPR) which improves speed by limiting the amount of graph data that need to be accessed.
BRPPR iteratively divides the vertices of a graph into an active and an inactive set.
At each iteration, the set of active vertices is expanded to include more vertices that are likely to have a high RWR score.
BRPPR expands nodes until the total rank on the frontier set of nodes is less than $\kappa$.
Proposed by Tong et al.~\cite{tong2008random}, NB-LIN exploits linear correlations across rows and columns of the adjacency matrix in many real-world graphs.
NB-LIN computes low-rank approximation of the adjacency matrix and uses it to estimate RWR score vector based on the Sherman-Morrison lemma.
NB-LIN divides whole computation into the preprocessing phase and online phase, and yields faster response time in the online phase.
Shin et al. extended their exact RWR method BEAR~\cite{DBLP:journals/tods/JungSSK16, shin2015bear} to an approximate RWR method BEAR-APPROX which drops non-zero entries whose absolute value is smaller than the drop tolerance in its preprocessed matrix.
Forward Push~\cite{andersen2006local} computes RWR by propagating residuals across a graph until all the residuals become smaller than a given threshold.
Proposed by Wang et al.~\cite{wang2017fora}, FORA first performs Forward Push with early termination, and subsequently runs random walks.
FORA utilizes Forward Push to significantly cut down the number of required random walks while satisfying the same result quality guarantees of random walks.
FORA precomputes a number of random walks in the preprocessing phase to further improve computation efficiency.
Other methods such as FAST-PPR~\cite{lofgren2014fast}, BiPPR~\cite{lofgren2016personalized} and HubPPR~\cite{wang2016hubppr} narrow down the scope of RWR problem by specifying a target node.
BiPPR processes an RWR query through a bi-directional search on the input graph.
HubPPR precomputes indexes in the preprocessing phase and approximates RWR with the help of precomputed indexes in the bi-directional way.
We compare our method with HubPPR since HubPPR is the most recent study with the best performance among bi-directional methods~\cite{wang2016hubppr}.
Our proposed \methodA outperforms all methods described above by providing a better cost-efficiency. 


\vspace{10pt}
\section{Conclusion}
\label{sec:conclusion}
\begin{figure*}[t!]
	\centering
	\vspace{-3mm}
	\includegraphics[width=.05\linewidth]{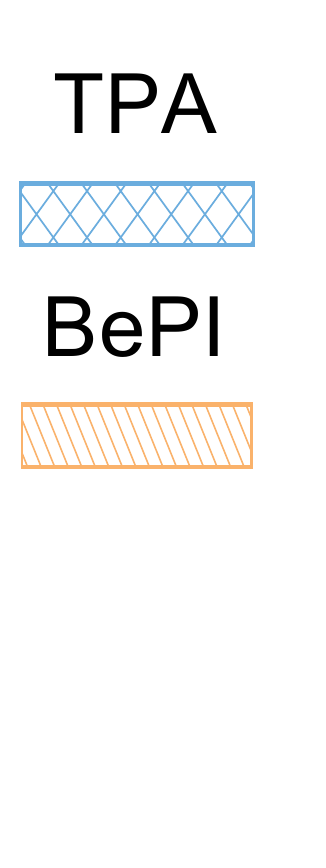}
	\hspace{-3mm}
	\subfigure[Size of preprocessed data]
	{
		\label{fig:bepi:memory}
		\includegraphics[width=.3\linewidth]{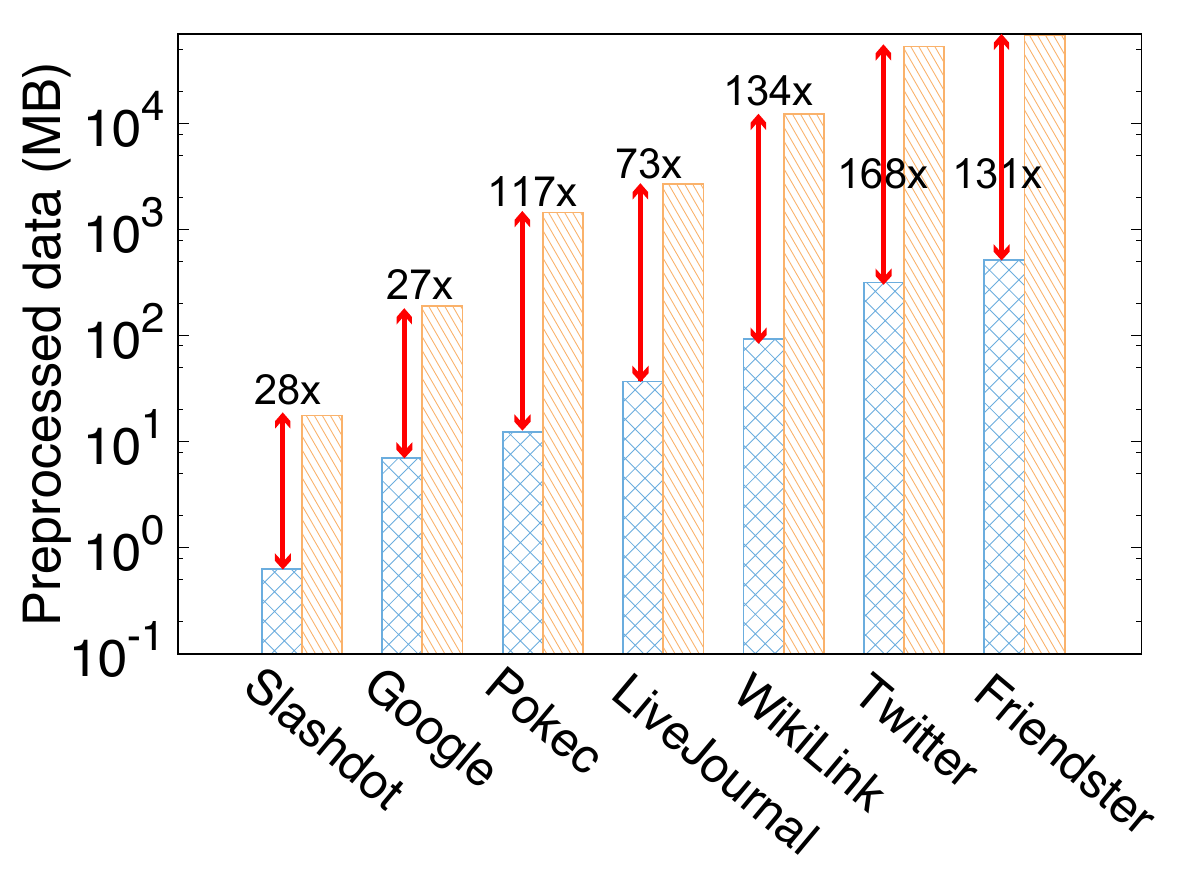}	
	}
	\hspace{-3mm}
	\subfigure[Preprocessing time]
	{
		\label{fig:bepi:preprocess}
		\includegraphics[width=.3\linewidth]{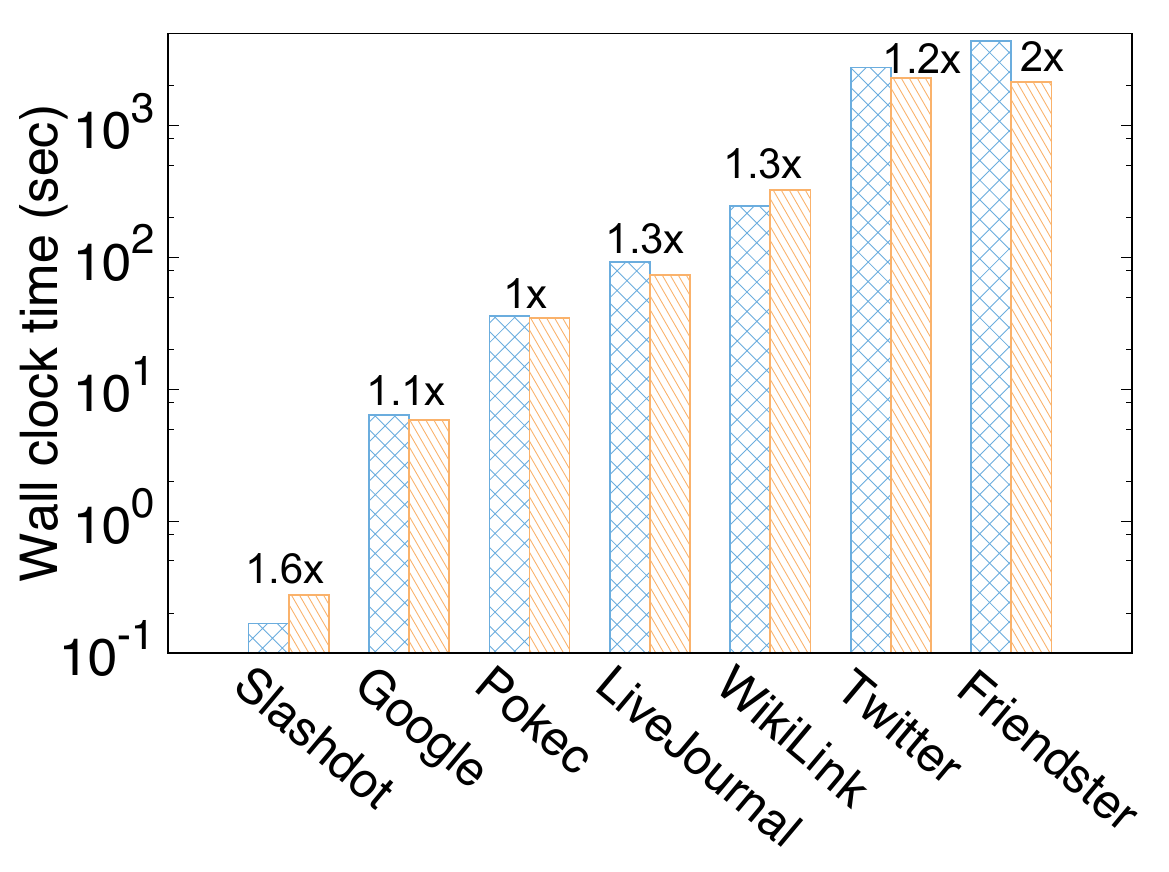}	
	}
	\hspace{-3mm}
	\subfigure[Online time]
	{
		\label{fig:bepi:online}
		\includegraphics[width=.3\linewidth]{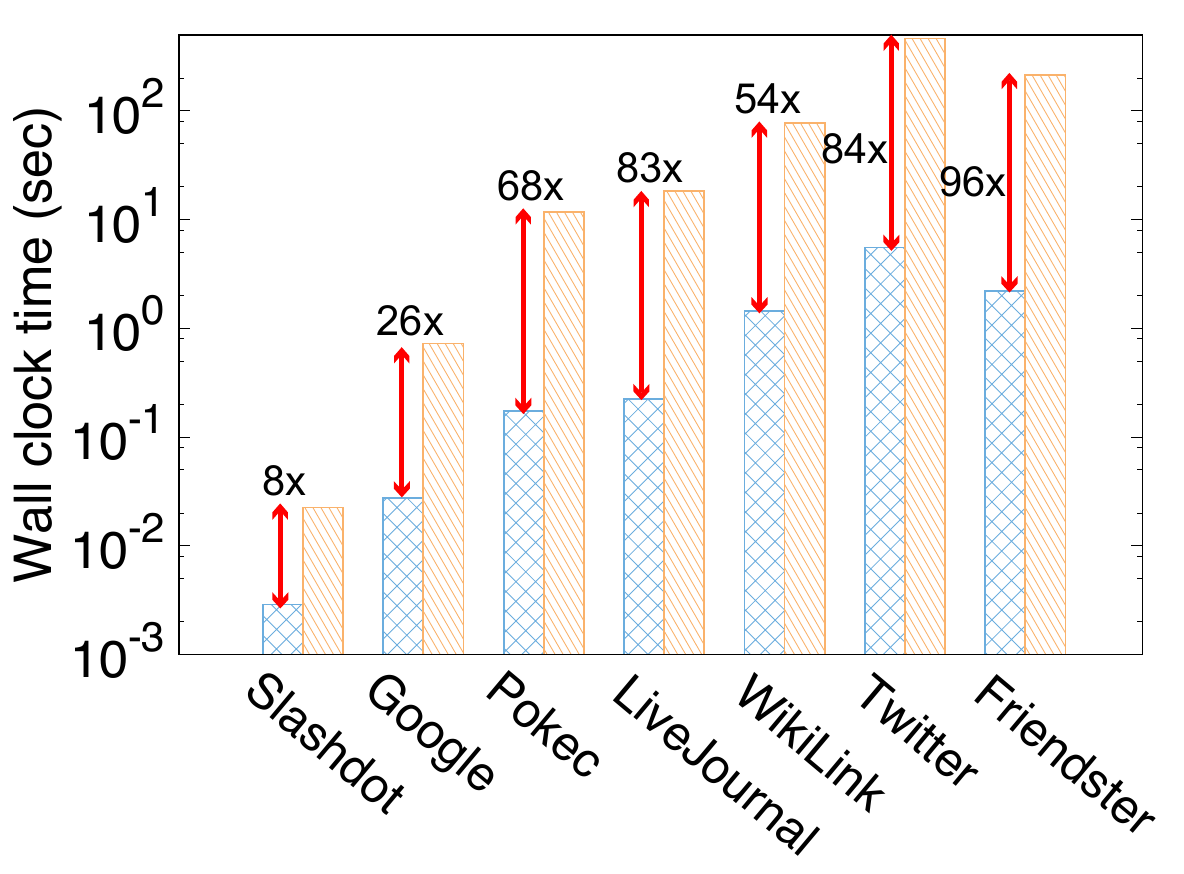}
	}
	\caption
	{ {Comparison with BePI:
			(a) \methodA uses less amount of space for preprocessed data than BePI does across all datasets.
			(b) In the preprocessing phase, \methodA and BePI take similar computation time.
			(c) In the online phase, \methodA computes RWR scores faster than BePI does over all datasets.
			Note that \methodA computes the approximate RWR scores while BePI results in the exact RWR scores.}
	}
	\label{fig:bepi}
\end{figure*}

In this paper, we propose TPA, a fast and accurate method for computing approximate RWR.
\methodA is based on  cumulative power iteration (\methodE) which interprets RWR problem as propagation of scores from a seed node across a graph.
To avoid long computation time, \methodA divides the whole iterations of \methodE into three parts (family, neighbor, and stranger parts), and estimates the neighbor part and the stranger part using our proposed approximation methods called neighbor approximation and stranger approximation, respectively. 
With the help of two approximation phases, \methodA quickly computes only the family part in the online phase, and then approximates RWR with high accuracy.
Our evaluation shows that \methodA outperforms other state-of-the-art methods in terms of speed and memory-efficiency, without sacrificing accuracy. 
Future works include extending \methodA into a disk-based RWR method to handle huge, disk-resident graphs.

\vspace{10pt}
\bibliography{BIB/myref}
\bibliographystyle{abbrv}

\vspace{10pt}
\appendix
\label{appendix}
\subsection{Comparison with BePI}
\label{appendix:bepi}

BePI~\cite{JungPSK17} is the state-of-the-art exact RWR method which precomputes several matrices required by the online phase in the preprocessing phase and computes RWR scores by exploiting the precomputed matrices in the online phase.
As shown in Figure~\ref{fig:bepi}, \methodA and BePI take the similar preprocessing time, while \methodA is up to $96\times$ faster than BePI in the online phase.
Considering that the preprocessing phase is executed only once for a graph and the online phase is executed everytime for a new seed node,
the superior performance of \methodA for online computation brings significant advantages for users who put more priority on speed than accuracy.
Moreover, \methodA requires up to $168\times$ less memore space for preprocessed data than BePI.
Note that while \methodA outperforms BePI in terms of computation time and memory usage,
\methodA computes the approximate RWR scores and BePI results in the exact RWR scores.


\end{document}